\documentclass[aps,a4paper,nobibnotes,floatfix,superscriptaddress,nofootinbib]{revtex4}

\usepackage{amssymb,amsmath,amsfonts}
\usepackage{graphicx}
\usepackage{bm}
\usepackage{paralist,array}
\usepackage{hyperref}

\newcommand{\be}{\begin{equation}}
\newcommand{\ee}{\end{equation}}
\newcommand{\bea}{\begin{eqnarray}}
\newcommand{\eea}{\end{eqnarray}}

\newcommand{\E}{{\mathbb E}}
\newcommand{\kmin}{k_\text{min}}
\newcommand{\qmin}{q_\text{min}}
\newcommand{\qmax}{q_\text{max}}
\newcommand{\qea}{q_\text{\tiny EA}}
\newcommand{\FRS}{F_\text{RS}}
\newcommand{\FRSB}{F_\text{RSB}}

\begin{document}

\title{Spin glasses in a field show a phase transition
varying the distance among real replicas (and how to exploit it to find the critical line in a field)}

\author{Maddalena Dilucca}
\affiliation{Dipartimento di Fisica,
Sapienza Universit\'a di Roma,
P.le A. Moro 2, I-00185 Roma, Italy}

\author{Luca Leuzzi}
\affiliation{CNR, Nanotec, Rome unit,
P.le A. Moro 2, I-00185 Roma, Italy}
\affiliation{Dipartimento di Fisica,
Sapienza Universit\'a di Roma,
P.le A. Moro 2, I-00185 Roma, Italy}

\author{Giorgio Parisi}
\affiliation{Dipartimento di Fisica,
Sapienza Universit\'a di Roma,
P.le A. Moro 2, I-00185 Roma, Italy}
\affiliation{CNR, Nanotec, Rome unit,
P.le A. Moro 2, I-00185 Roma, Italy}
\affiliation{INFN, Sezione di Roma I,
P.le A. Moro 2, I-00185 Roma, Italy}

\author{Federico Ricci-Tersenghi}
\affiliation{Dipartimento di Fisica,
Sapienza Universit\'a di Roma,
P.le A. Moro 2, I-00185 Roma, Italy}
\affiliation{CNR, Nanotec, Rome unit,
P.le A. Moro 2, I-00185 Roma, Italy}
\affiliation{INFN, Sezione di Roma I,
P.le A. Moro 2, I-00185 Roma, Italy}

\author{Juan J. Ruiz-Lorenzo}
\affiliation{Departamendo de F\'{\i}sica and ICCAEx,
Universidad de Extremadura,
06006 Badajoz, Spain}
\affiliation{Instituto de Biocomputaci\'on y F\'{\i}sica de los Sistemas Complejos (BIFI), 50018 Zaragoza, Spain}

\date{\today}

\begin{abstract}
We discuss a phase transition in spin glass models which have been
rarely considered in the past, namely the phase transition that may
take place when two real replicas are forced to be at a larger
distance (i.e.\ at a smaller overlap) than the typical one.

In the first part of the work, by solving analytically the
Sherrington-Kirkpatrick model in a field close to its critical point,
we show that even in a paramagnetic phase the forcing of two real
replicas to an overlap small enough leads the model to a phase
transition where the symmetry between replicas is spontaneously
broken. More importantly this phase transition is related to the de
Almeida-Thouless (dAT) critical line.

In the second part of the work, we exploit the phase transition in the
overlap between two real replicas to identify the critical line in a
field in finite dimensional spin glasses. This is a notoriously
difficult computational problem, because of huge finite size
corrections.  We introduce a new method of analysis of Monte Carlo
data for disordered systems, where the overlap between two real
replicas is used as a conditioning variate.  We apply this analysis to
equilibrium measurements collected in the paramagnetic phase in a
field, $h>0$ and $T_c(h) < T < T_c(h=0)$, of the $d=1$ spin glass
model with long range interactions decaying fast enough to be outside
the regime of validity of the mean-field theory.  We thus provide very
reliable estimates for the thermodynamic critical temperature in a
field.
\end{abstract}

\maketitle

\section{Introduction}

The study of spin glass models in an external field started more than 40 years ago \cite{dealmeida:78}, but has demonstrated to be an extremely challenging problem. The results of the many numerical simulations performed over the last three decades \cite{caracciolo:90,huse:91,caracciolo:91,ciria:93,parisi:98b,marinari:98e,marinari:98g,marinari:98h,houdayer:99,marinari:00d,houdayer:00,cruz:03,young:04,leuzzi:08,leuzzi:09,leuzzi:11,janus:12,larson:13,janus:14b,janus:14c} and
have been non conclusive and often interpreted in contradicting ways.
The main reason for this difficulty seems to rely in the huge finite size corrections that spin glasses have in presence of an external magnetic field, that in turn make very difficult to extract the thermodynamic behavior.

The presence of strong finite size corrections was known since the very first numerical simulations \cite{caracciolo:90,huse:91,caracciolo:91,ciria:93}. However, only recently it has been possible to obtain a quantitative measure of it by running Monte Carlo simulations in a model whose critical behavior is known analytically \cite{takahashi2010finite}. Indeed, by using the cavity method \cite{mezard:01}, spin glass models on random graphs can be analytically solved in the paramagnetic phase and the critical dAT line separating the paramagnetic and the spin glass phases can be computed exactly. However a standard analysis of Monte Carlo measurements taken from a spin glass model in a field defined on a random graph fail to identify correctly the critical temperature \cite{takahashi2010finite}. A possible explanation of this surprising finding comes from the observation made in Ref.~\cite{parisi2012numerical} that mean values of the observables used in the Monte Carlo analysis are dominated by a minority of atypical measurements (e.g.\ atypical in the value of the mean overlap $q$) and a better analysis focusing on typical measurements can identify correctly the critical point predicted analytically. 

A similar problem in the analysis of Monte Carlo data measured from a spin glass model in a field has been found in Ref.~\cite{leuzzi:09}, where it was found that a standard finite size scaling analysis was unable to identify the correct critical point in presence of a field, while the same analysis works perfectly in the absence of the external field.
The main source of fluctuations, leading to dominance of atypical measurements, was identified in the value of the 4-points correlation at large distance, i.e.\ $q^2$, where $q=\sum_i s_i t_i /N$ is the overlap between the two  simulated real replicas ${\bm s}$ and ${\bm t}$. This value enters in the Fourier transform of the correlation function at zero wavelength ($k=0$).
By performing a new analysis that avoids the use of the $k=0$ component, and restricts to $k\in\{2\pi/L,4\pi/L\}$, it was possible to clearly identify a phase transition in a field \cite{leuzzi:09}.

Subsequent works \cite{leuzzi:11,janus:12,janus:14c} confirmed with more statistics and in a broader class of spin glass models in a field that the main source of these huge finite size corrections is in the behavior of atypical samples and/or atypical measurements.
In summary, what seems to happen in spin glass models under the effect of an external field is the following.
Even for the largest systems that can be simulated with present computer facilities the probability distribution of the overlap is very broad and shows an exponential tail extending in the region of small and even negative overlaps. This tail is clearly a finite size effect, that must disappear in the
thermodynamic limit. However, the measurements corresponding to these atypically small values of the overlap tend to dominate the average in samples of finite size and to hide the behavior of the vast majority of measurements.

In order to extract the typical behavior, in Ref.~\cite{janus:14c} all
the measurements have been divided in ten deciles according to a
conditioning variate, i.e., essentially the overlap among the
simulated replicas. The results clearly show that the behavior of the
first (or the last) decile is definitely different from the median
behavior. Moreover, the behavior of the median decile is the one
showing the least finite size effects and, therefore, it is likely to
approach the thermodynamic behavior faster in $N$. From measurements
based on data relative to this median decile in $q$ one can observe a
clear phase transition that was impossible to identify taking the
unconditional average on all the measurements.

So the general picture that emerges from several different analysis carried out until now is that the behavior of spin glasses in a field may strongly depends on the specific value of the overlap between the real replicas simulated. And this seems to be true even in the paramagnetic phase, $T>T_c(h)$, where asymptotically in the large $N$ limit the overlap takes a unique value, $q=\qea$, with high probability.

Therefore, it is seems  natural to us to consider a spin glass model in a field with two real replicas constrained to take an overlap equal to $q$. And check how much the behavior of the model depends on the value of $q$.
Moreover, fixing the overlap between the two real replicas to $q$, one has the advantage that fluctuations in $q$ are suppressed and this might only produce a better signal-to-noise ratio in the analysis of numerical data.

In the following we provide detailed information on the above points. In Section~\ref{sec:MF} we solve a mean-field model for spin glasses in a field and show indeed that a phase transition takes place if the overlap between the real replicas is made small enough, and this can be connected to the critical dAT line.
Then, in Section~\ref{sec:numerics}, we analyze numerical data for a finite-dimensional spin glass in a field conditioning on the overlap between the two simulated replicas, finding, indeed, a phase transition towards a spin glass phase. This conditioned analysis provides reliable estimates for the location of the critical dAT line $T_c(h)$.

\section{Phase transition varying the overlap between two real replicas in a solvable mean-field model}
\label{sec:MF}

Our interest is in understanding what happens to the paramagnetic phase of a disordered model when two real replicas are coupled and forced to be at an atypical overlap value. In order to make contact with previous literature \cite{franz:95}, in this Section we call $p_d$ the overlap at which the two real replicas are forced to be. In particular, we would like to uncover possible phase transitions upon changing $p_d$ and connect these with the  phase transition on the dAT line.

In order to work with a solvable mean-field model we consider the so-called truncated model that describes correctly the Sherrington-Kirkpatrick (SK) model close to its critical point $(T=T_c=1,h=0)$. The equations of two replicas constrained to be at overlap $p_d$ were already considered in Ref.~\cite{franz1992replica}. However in that work the authors focused mainly on the Replica Symmetry Breaking (RSB) solutions, while we are mostly interested in identifying eventual phase transitions taking place in the paramagnetic phase when $p_d<\qea$.

\subsection{The truncated model}

The truncated model is an expansion of the free-energy of the SK model in powers of the $Q$ matrix up to the fourth order term, which is responsible for the breaking of the replica symmetry \cite{parisi1980order}.
Under this ``truncated'' approximation, the free-energy reads
\begin{equation}
F= \tau \langle q^2 \rangle - \frac{1}{3} \left[2\langle q \rangle \langle q^2 \rangle+\int_{0}^{1} dx\,q(x) \int_{0}^{x} dz\,[q(x)-q(z)]^2 \right]+  \frac{1}{4} y \langle q^4 \rangle +h^2 \langle q \rangle\,,
\end{equation}
where $\langle q^k \rangle \equiv \int_0^1 q(x)^k dx$, being $q(x)$ the Parisi order parameter, $\tau=1-T/T_c(h=0)=1-T$ and $y=2/3$ for the SK model.
The function $q(x)$ extremizing the truncated free-energy has been computed in Ref.~\cite{parisi1980order} and reads
\bea
q(x) =& \qea(\tau,h) \phantom{XXXXXXXXXXXX} & \text{for }\tau \le \tau_c(h)\,,\\
q(x) =& \left\{
\begin{array}{ll}
\qmin(h,y) & 0 \le x \le 3y \qmin \,,\\
\frac{x}{3y} & 3y\qmin < x < 3y\qmax\,,\\
\qmax(\tau,y) & 3y \qmax \le x \le 1\,,
\end{array}
\right.
& \text{for }\tau > \tau_c(h)\,,
\eea
where
\be
\qmin(h,y) = \left(\frac{h^2}{2y}\right)^{1/3}\,, \quad \qmax(\tau,y) = \frac{1 - \sqrt{1 - 6 y \tau}}{3 y}\,,
\label{eq:qmin-qmax}
\ee
and $\qea(\tau,h,y)$ satisfies
\be
h^2+2\tau\qea-2\qea^2+y\qea^3=0\,.
\label{eq:qea}
\ee

The dAT line signaling the onset of the RSB phase can be obtained by imposing the condition $\qmin(h,y)=\qmax(\tau,y)$ and, to leading order, is given by $\tau_c(h,y) = (h^2/(2y))^{1/3}$ or $h_c(\tau,y) = \sqrt{2y}\,\tau^{3/2}$.
It is worth noticing that in the paramagnetic phase, i.e.\ for $\tau < \tau_c(h,y)$ or $h > h_c(\tau,y)$, the order of the three overlaps just defined is $\qmax(\tau,y) < \qea(\tau,h,y) < \qmin(h,y)$.

\subsection{The model with constrained replicas}

The case of two real replicas constrained to have overlap $p_d$ has been considered in Ref.~\cite{franz1992replica}. A more complex order parameter that involves two matrices $Q$ and $P$ is required, where $Q$ (resp.~$P$) describes the overlaps between copies of the same (resp.\ different) real replica(s). Matrices $Q$ and $P$ are parametrized as usual via the Parisi functions $q(x)$ and $p(x)$, thus getting the following free-energy
\begin{align}
F = & \tau \left[\langle q^2 \rangle + \langle p^2 \rangle -p_d^2 \right] -  \frac{1}{3} \bigg[ 2 \langle q \rangle \langle q^2 \rangle+  \int_0^1 dx\,q(x) \int_0^x dz\,[q(x)-q(z)]^2 + 6\langle pq \rangle (\langle p \rangle -p_d) \nonumber\\
& + 3 \int_0^1 dx\,q(x) \int_0^x dz\,[p(x)-p(z)]^2 \bigg]  + \frac{y}{4} [\langle q^4 \rangle +\langle p^4 \rangle -p_d^4]+  h^2[\langle q \rangle +\langle p \rangle -p_d]\,.
\end{align}
We consider $p_d$ as a free parameter that we want to change in order to test for eventual phase transitions when $p_d$ becomes small. So the free-energy needs to be extremized only with respect to $q(x)$ and $p(x)$. The corresponding equations are the following
\begin{equation}
\label{RSBspe}
\begin{split}
\frac{\delta F}{\delta q(x)} &= 2(\tau-\langle q \rangle)q(x)+2(p_d-\langle p \rangle)p(x)-\int_0^x dz\,[q(x)-q(z)]^2 - \int_0^x dz\,[p(x)-p(z)]^2 + yq^3(x)+h^2=0\,,\\
\frac{\delta F}{\delta p(x)} &= 2(\tau-\langle q \rangle)p(x)+2(p_d-\langle p \rangle)q(x)-2\int_0^x dz\,[q(x)-q(z)][p(x)-p(z)] +yp^3(x)+h^2=0\,.
\end{split}
\end{equation}
Taking some derivatives and doing a little bit of algebra one can prove in general the following statements~\cite{franz1992replica}:
\begin{itemize}
\item if $q'(x)=0$, then $p'(x)=0$.
\item If $q'(x)\neq 0$ and $p'(x)=0$, then $q(x)=\frac{x}{3y}$.
\item if $p(z)=q(z)\;\forall z\le x$, then either $q'(x)=0$ or $q(x)=\frac{2x}{3y}$.
\end{itemize}
We will investigate different quite general ansatz for $q(x)$ and $p(x)$ compatible with these conditions, but we do not find useful to write down the most general ansatz compatible with the constraints because it is rather cumbersome.
We prefer to add complexity to the solution step by step.

\subsection{Replica Symmetry (RS) solutions}

We start with the RS solution $q(x)=q$ and $p(x)=p$, that certainly holds above the dAT line, when the constraining overlap $p_d$ is close to the typical value for the overlap $\qea(\tau,h,y)$.
The equations to be solved are
\bea
h^2 + 2 p_d q + 2 \tau p - 4 q p + y p^3 &=& 0 \label{eq:RS1}\,,\\
h^2 + 2 \tau q + 2 p_d p - 2 q^2 - 2 p^2 + y q^3 &=& 0 \,.\label{eq:RS2}
\eea
These equations admit a symmetric solution $p=q=\qea(p_d,\tau,h,y)$ with the latter defined by
\be
h^2 + 2 (\tau + p_d) \qea - 4 \qea^2 + y \qea^3=0\,.
\ee
The $\qea$ overlap in the model with two constrained replicas is related to the one in the model with a single replica through the following transformation of parameters: $\tau \to \frac{\tau+p_d}{2}, h\to\frac{h}{\sqrt{2}}, y\to\frac{y}{2}$. In other words
\be
\qea(p_d,\tau,h,y) = \qea\left(\frac{\tau+p_d}{2},\frac{h}{\sqrt{2}},\frac{y}{2}\right)\;.
\ee
From this observation it is easy to obtain the condition under which the RS symmetric solution $q=p=\qea(p_d,\tau,h,y)$ is stable with respect to a solution still symmetric, $q(x)=p(x)$, but breaking the replica symmetry:
\be
\qmax\left(\frac{\tau+p_d}{2},\frac{y}{2}\right) < \qmin\left(\frac{h}{\sqrt{2}},\frac{y}{2}\right)\;.
\label{eq:RS-RSBsymm}
\ee
This equation defines an upper bound on $p_d$, because $\qmax(\tau,y)$ is monotonously increasing in $\tau$.
Hereafter we will always work in the range of $p_d$ satisfying Eq.~(\ref{eq:RS-RSBsymm}).
For example for $\tau=0.1$, $h=0.1$, $y=2/3$ the bound reads $p_d<0.253171$ (these values for $\tau$, $h$ and $y$ define a point in the paramagnetic phase of the SK model and will be used as an example in the rest of this Section).
In the paramagnetic phase, fixing $p_d=\qea(\tau,h,y)$, that is constraining the real replicas to the typical value, the bound in Eq.~(\ref{eq:RS-RSBsymm}) is always satisfied.

\begin{figure}[t]
\centering\includegraphics[width=0.5\columnwidth]{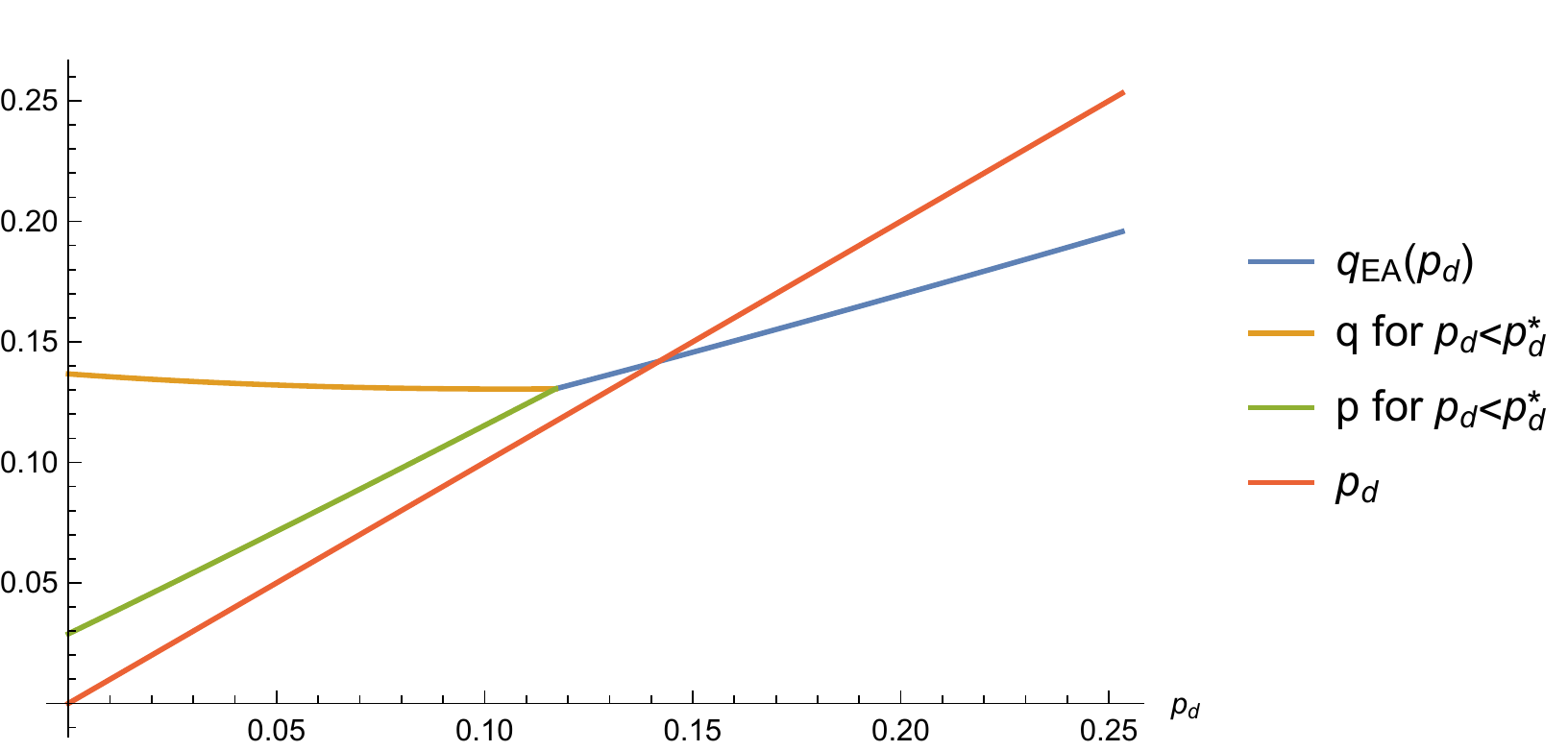}
\caption{Parameters of the RS solutions versus $p_d$ in the case of $\tau=h=0.1$ and $y=2/3$. The merging of the three curves takes place at $p_d=p_d^*(\tau,h,y)=0.117033$, while the crossing between the two curves takes place at $\qea(\tau,h,y)=0.141942$.}
\label{fig:RSoverlaps}
\end{figure}

Lowering enough the value of $p_d$, the symmetry $p(x)=q(x)$ can spontaneously break down. At the RS level this actually corresponds to the free-energy being extremized by a solution with $p\neq q$. Such a phase transition takes place at $p_d=p_d^*$ where $p_d^*(\tau,h,y)$ can be obtained from the linearization of Eqs.~(\ref{eq:RS1}-\ref{eq:RS2}) around the solution $p=q=\qea(p_d,\tau,h,y)$ and solves the following equation
\be
p_d^* = \tau + \frac{3y}{2} \big(\qea(p_d^*,\tau,h,y)\big)^2\;.
\label{eq:pdstar}
\ee
Then for $p_d \ge p_d^*$ we have $p=q=\qea(p_d,\tau,h,y)$, while for $p_d < p_d^*$ we have $p<q$. See Fig.~\ref{fig:RSoverlaps} for an example in the case $\tau=h=0.1$ and $y=2/3$.

\begin{figure}[t]
\centering\includegraphics[width=0.55\columnwidth]{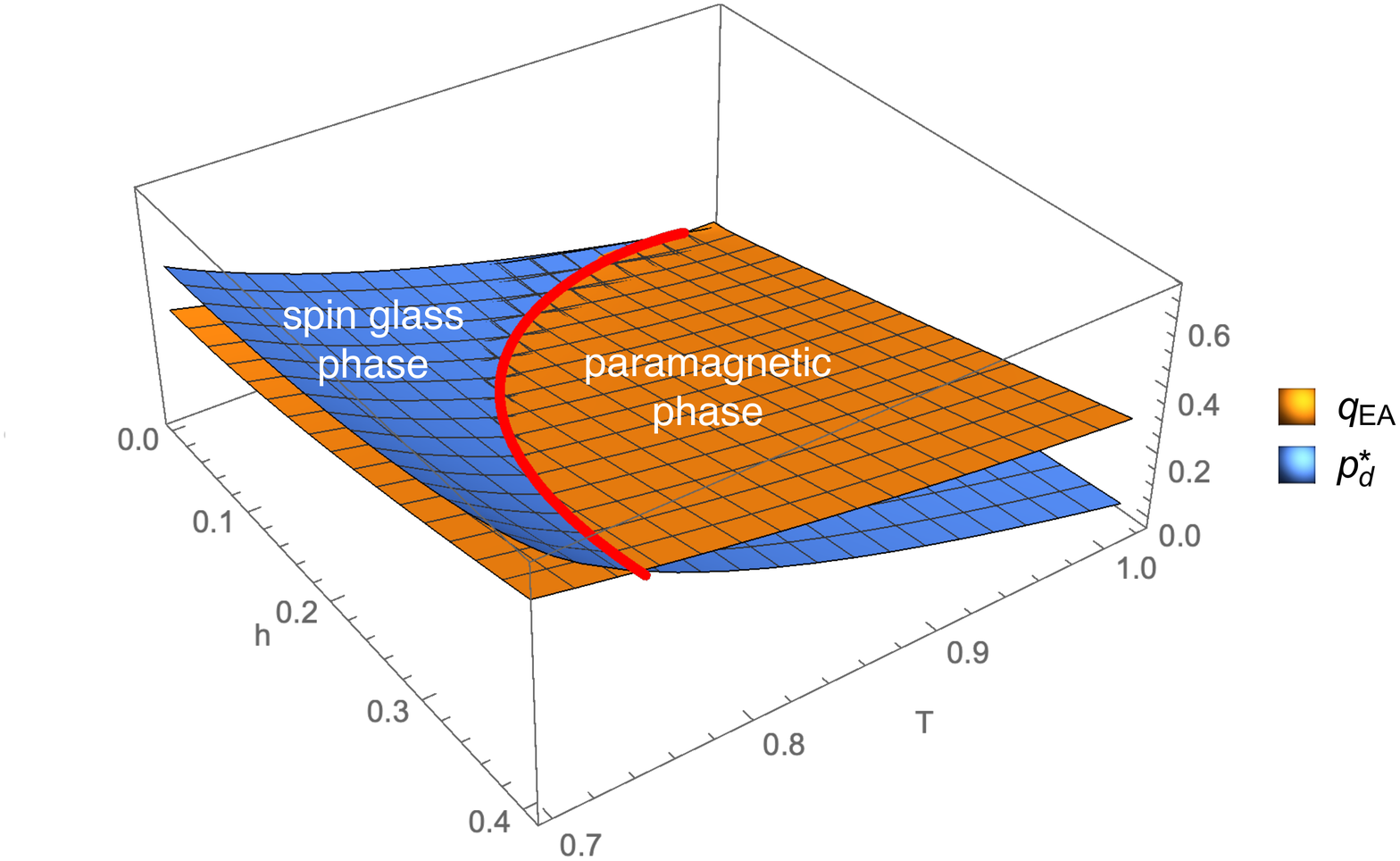}
\caption{Values of $\qea(\tau,h,y)$ and $p_d^*(\tau,h,y)$ plotted in the $(h,T=1-\tau)$ plane for $y=2/3$. The red bold curve is the dAT line, separating the paramagnetic and the spin glass phases. $\qea$ and $p_d^*$ merge on the dAT line, while their values in the spin glass phase have no physical meaning. Below the blue surface in the paramagnetic phase the $p(x)=q(x)$ symmetry is broken.}
\label{fig:qea_pd}
\end{figure}

The inequality $p_d^*(\tau,h,y)<\qea(\tau,h,y)$ that we can easily check in Fig.~\ref{fig:RSoverlaps} for $\tau=h=0.1$ and $y=2/3$ is a general feature of the paramagnetic phase.
In Fig.~\ref{fig:qea_pd} we show $\qea(\tau,h,y)$ and $p_d^*(\tau,h,y)$ for $y=2/3$ and we notice that the dAT line separating the paramagnetic and spin glass phases corresponds exactly to the locus where $\qea$ and $p_d^*$ coincide. Below the blue surface in the paramagnetic phase the $p(x)=q(x)$ symmetry is broken.

\begin{figure}
\centering\includegraphics[width=0.54\columnwidth]{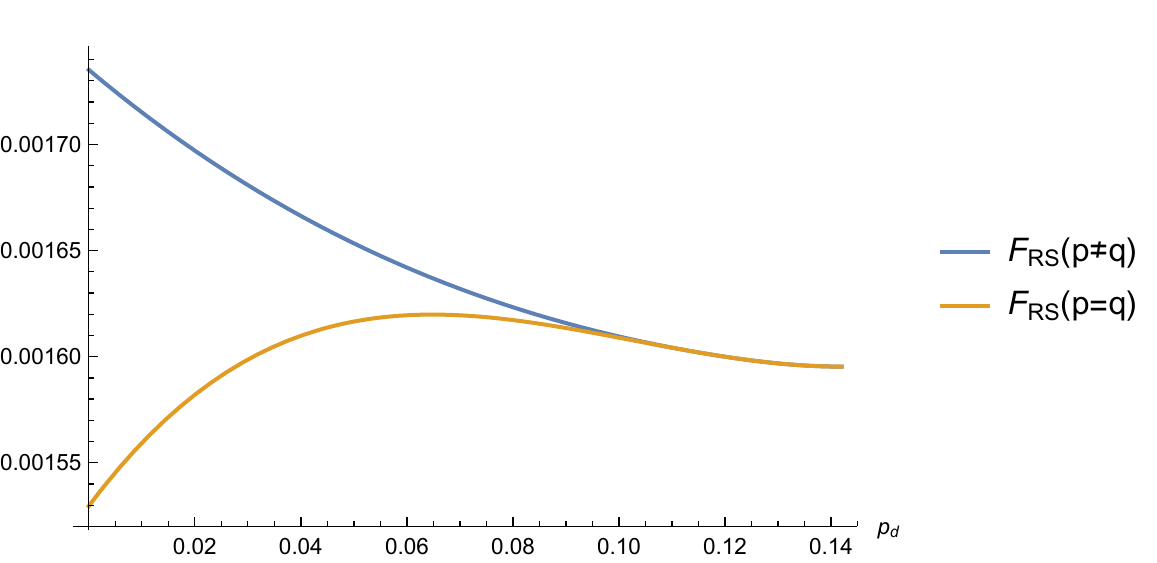}
\centering\includegraphics[width=0.45\columnwidth]{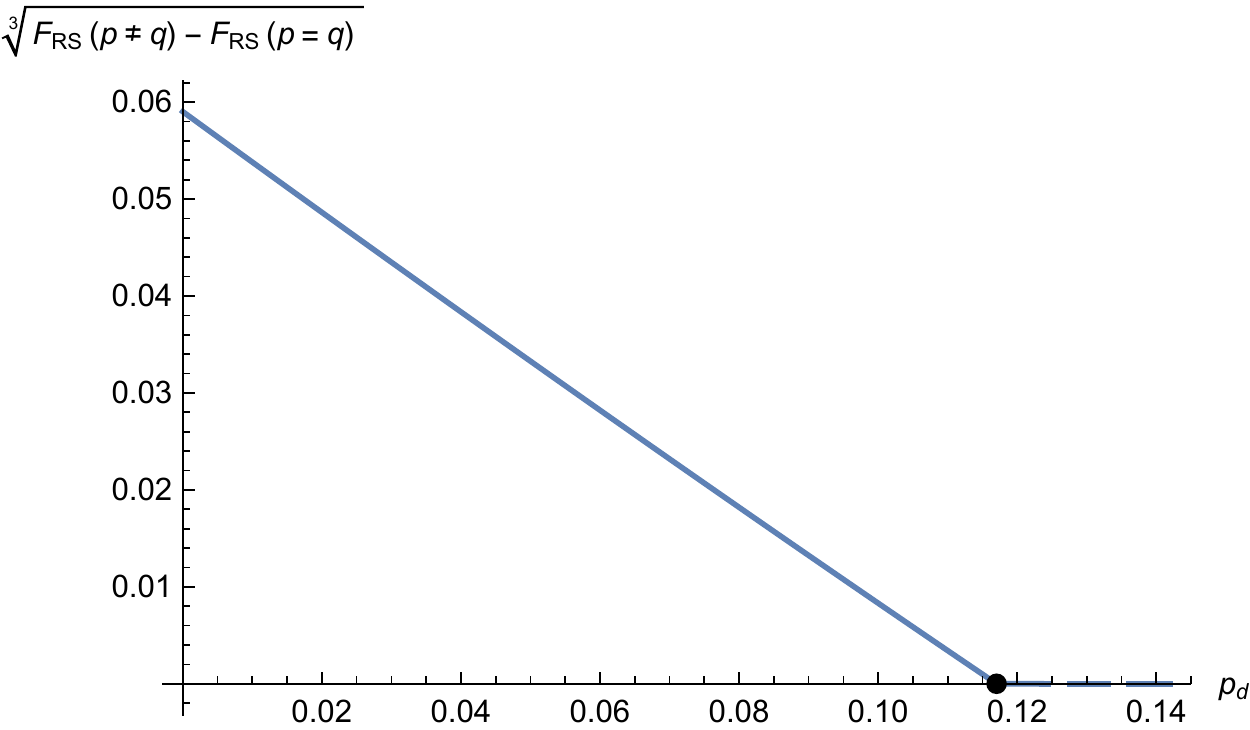}
\caption{Free-energies of the $p=q$ and $p\neq q$ RS solutions for $\tau=h=0.1$ and $y=2/3$. Below $p_d^*=0.117033$ the free-energy of the $p\neq q$ solution is higher and such a solution dominates over the symmetric one (left panel). The free-energy difference goes like $(p_d^*-p_d)^3$ as can be seen in the right panel, where the black dot marks the value of $p_d^*$.}
\label{fig:FRS}
\end{figure}

In Fig.~\ref{fig:FRS} we show for $\tau=h=0.1$ and $y=2/3$ the free-energies of the $p=q$ and $p\neq q$ solutions. Below $p_d^*=0.117033$ the free-energy of the $p\neq q$ solution is higher and such a solution dominates over the symmetric one (left panel). The free-energy difference goes like $(p_d^*-p_d)^3$ as can be seen from the right panel, where the black dot marks the value of $p_d^*$.

\subsection{Replica Symmetry Breaking (RSB) solutions in the paramagnetic phase}

The next step is to show that at $p_d^*$, where the $p(x)=q(x)$ symmetry gets spontaneously broken, also the replica symmetry spontaneously breaks. In order to show this we have to search for solutions to the saddle point equations in Eq.~(\ref{RSBspe}) with a RSB order parameter.
We assume a single breaking point for both $p(x)$ and $q(x)$, that is we use the following ansatz
\be
p(x) = \left\{
\begin{array}{ll}
p_0 & 0 \le x < m\\
p_1 & m < x \le 1
\end{array}
\right.
\qquad
q(x) = \left\{
\begin{array}{ll}
q_0 & 0 \le x < m\\
q_1 & m < x \le 1
\end{array}
\right.
\ee
We have now five  saddle point equations to fix $p_0,p_1,q_0,q_1$ and $m$. We have looked numerically to their solutions and we have found that, beyond the RS solution $p_0=p_1=p$ and $q_0=q_1=q$, other two solutions exist:
\begin{enumerate}
\item a solution with $p_0 \gtrsim p_1 \simeq p$ and $q_0 \lesssim q_1 \simeq q$, that is with the $p(x)$ and $q(x)$ respectively very close to the RS corresponding overlaps $p$ and $q$,
\item a solution with $p_1 \simeq p < p_0 = q_0 < q_1 \simeq q$, that is where $p_1$ and $q_1$ are close to the RS overlaps and at small $x$ roughly a mean overlap is found $p_0=q_0 \simeq \frac{p+q}{2}$.
\end{enumerate}
We observe that in both these solution $p(x)$ is a non-increasing function, since $p_0>p_1$. This may seem at odd with the standard interpretation of the hierarchical structure of states in the SK model, where $q(x)$ is required to be a non-decreasing function.
However what is required for a correct physical interpretation of the states is the positivity of the matrix $\left(\begin{array}{cc}
Q & P\\
P & Q
\end{array}\right)$,
which can be ensured if a decrease in the function $p(x)$ is compensated by a larger increase in the function $q(x)$, i.e.\ if $q_1-q_0 > p_0-p_1$. We have checked that all the solutions found satisfy this criterion.

\begin{figure}
\centering\includegraphics[width=0.7\columnwidth]{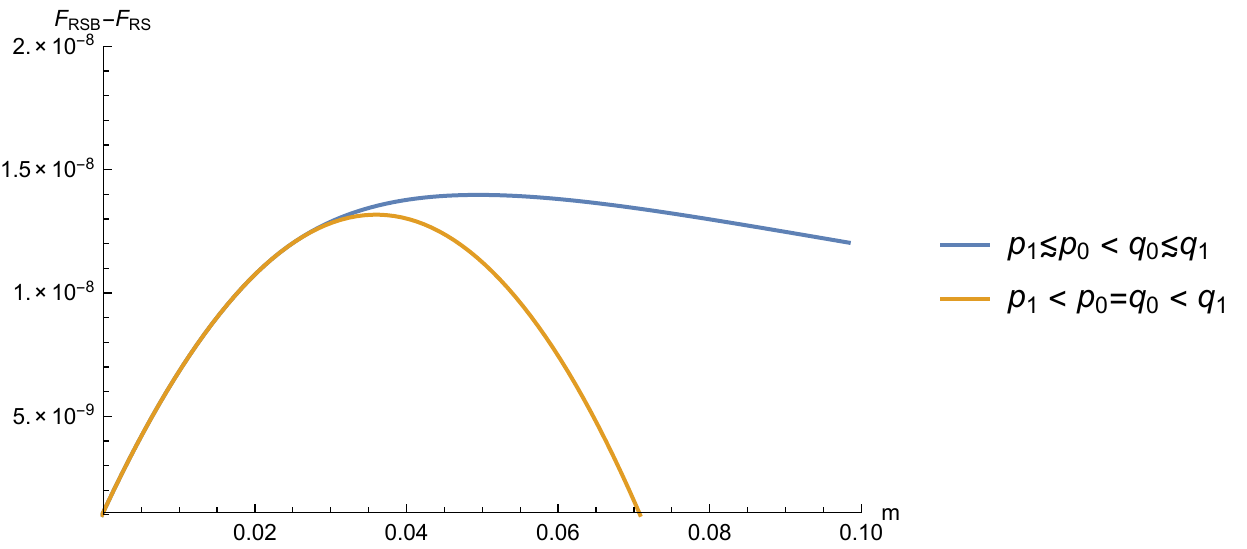}
\caption{Difference between the 1RSB free-energies and $F_\text{RS}$ for the two solutions with $\tau=h=0.1$, $y=2/3$ and $p_d=0.05$. We notice that the difference is very small, but clearly non-zero. Moreover the maximum is achieved for a rather small value of m, thus limiting the difference with respect to the RS solution to very small values of $x$ (remind that in both 1RSB solutions $p_1\simeq p$ and $q_1 \simeq q$).}
\label{fig:Fdiff_manysol}
\end{figure}

We have found that these one-step RSB (1RSB) solutions have a slightly better free-energy with respect to the RS solution with $p\neq q$.
In Fig.~\ref{fig:Fdiff_manysol} we show for $\tau=h=0.1$, $y=2/3$ and $p_d=0.05$ the difference between the 1RSB free-energies and $F_\text{RS}$ for the two solutions listed above. We notice that the difference is very small, but clearly non-zero. Moreover the maximum is achieved for a rather small value of $m$, thus limiting the difference with respect to the RS solution to very small values of $x$ (remember that in both the 1RSB solutions $p_1\simeq p$ and $q_1 \simeq q$).

\begin{figure}
\includegraphics[width=\columnwidth]{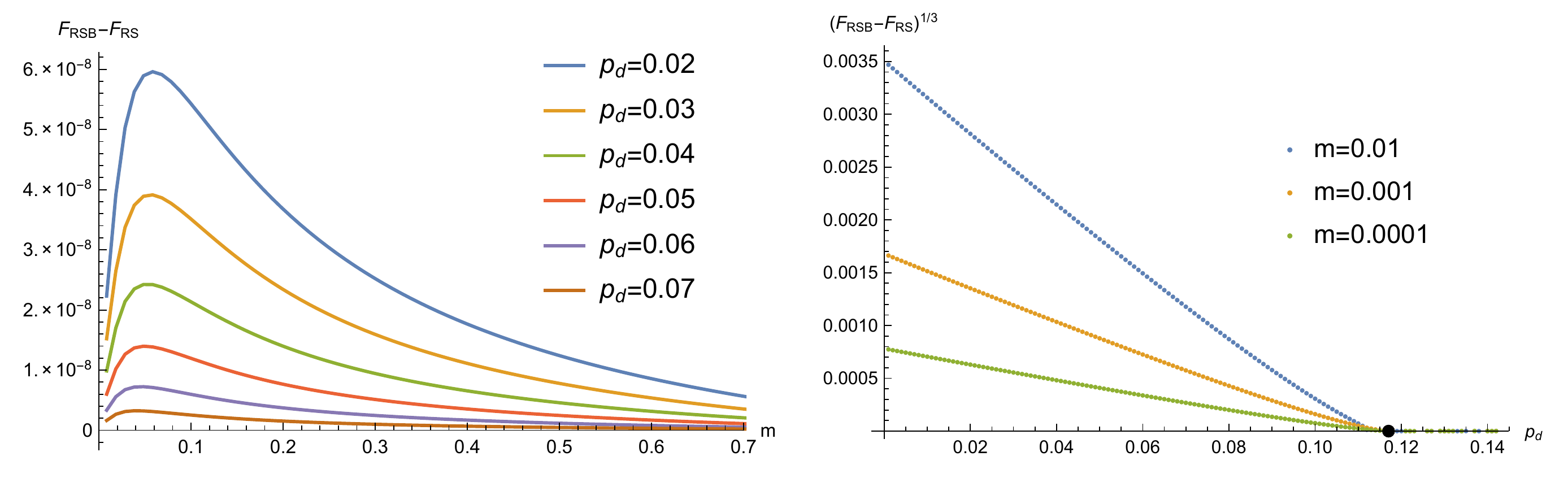}
\caption{Difference between the dominating 1RSB free-energy and $F_\text{RS}$ as a function of $m$ (left) and $p_d$ (right). The left panel shows that the location of the maximum of $F_\text{RSB}$ slightly decreases when $p_d$ grows, but the main effect is that for any $m$ value $F_\text{RSB}$ tends to $F_\text{RS}$ when $p_d$ grows. The right panel shows that for different $m$ values the free-energy difference goes to zero very close to $p_d^*$, marked with a black dot. Please notice that data in the region close to $p_d^*$ may have some uncertainty due to the extremely small free-energies difference, which are of the order $O(10^{-12})$.}
\label{fig:Fdiff}
\end{figure}

Although the free-energy difference between the two RSB solutions found is extremely small, the one to be preferred is the first one, where $p(x) \simeq p$ and $q(x) \simeq q$ (see Fig.~\ref{fig:Fdiff_manysol}). We have studied how the free-energy difference between this solution and the RS one changes varying the value of $p_d$. The results are shown in the left panel of Fig.~\ref{fig:Fdiff}: the location of the maximum of $F_\text{RSB}$ slightly decreases when $p_d$ grows, but the main effect is that for any $m$ value $F_\text{RSB}$ tends to $F_\text{RS}$ when $p_d\to p_d^*$ from below.

In order to better estimate the $p_d$ value where $\FRSB$ and $\FRS$ become equal we have plotted in the right panel of Fig.~\ref{fig:Fdiff} $(\FRSB-\FRS)^{1/3}$ as a function of $p_d$. This choice is dictated by the expectation that free-energy differences are cubic as in the RS case above and in previous studies \cite{franz1992replica}. Since the location of the maximum in the left panel changes a little bit with $p_d$, we show in right panel three different values of $m$. We observe that all the three curves extrapolate linearly to zero at a $p_d$ value very close to $p_d^*$, marked with a black dot. Please notice that data in the region close to $p_d^*$ may have some uncertainty due to the extremely small free-energies difference, which are of the order $O(10^{-12})$.

We have verified that the same analysis also holds for other values of $\tau$ and $h$. We then conclude that in the paramagnetic phase of the SK model, lowering the relative overlap between two real replicas the system undergoes at $p_d^*$ a phase transition, where the main effect is the breaking of the $p(x)=q(x)$ symmetry. Additionally at the same critical point, or very close to it, also the replica symmetry gets broken.

The identification of the location $p_d^*$ of this phase transition in the paramagnetic phase is very useful in order to locate the dAT line given that the latter coincide with the condition $p_d^*=\qea$ which can be checked in numerical simulations. This will be our aim in the next Section.

\section{Numerical results in a finite-dimensional spin glass model varying the overlap between two real replicas}
\label{sec:numerics}

According to the analytical results derived in the previous Section a spin glass model in a paramagnetic phase with a non-zero field, i.e. with $h>0$ and $T_c(h) < T<T_c(h=0)$, is expected to undergo a phase transition to a spin glass phase as the overlap $p_d$ between two real replicas is decreased to a value $p_d^*$. Moreover, on the dAT line the equality $p_d^*=\qea$ holds.

In this Section we analyse the numerical data collected in the paramagnetic phase of a spin glass model with a non-zero field with the aim of identifying $p_d^*$ and $\qea$. Given that we prefer to keep a different notation for the analytical critical point $p_d^*$ and the numerical estimate of the phase transition in $q$, we call the latter $q_c$ (but the reader should keep in mind that $q_c$ is the best numerical estimate for $p_d^*$).

\subsection{Model and numerical simulations}

Being interested in finite-dimensional spin glass models in a magnetic field, whose Hamiltonian is given by
\be
\mathcal{H}(\boldsymbol{s}) = -\sum_{i,j} J_{ij} s_i s_j - h \sum_i s_i\;,
\ee
we study a $d=1$ diluted spin glass model with long range interactions introduced in Ref.~\cite{leuzzi:08}. In this model each spin interacts on average with six neighbors and the interactions are present with a probability depending on the distance, i.e.\ $J_{ij} = \pm1$ with probability $\mathbb{P}[J_{ij} \neq 0] \propto |i-j|^{-\rho}$. Changing the exponent $\rho$ the effective dimension of the model varies \cite{leuzzi:08} and for $\rho>\rho_U=4/3$ the model is outside the range of validity of the mean-field theory and thus presents a non trivial critical behavior (as in a finite-dimensional spin glass model). The advantage of this model, with respect to the fully-connected version \cite{kotliar:83,leuzzi:99} is that the finite connectivity allows to simulate very large sizes, up to $L \sim O(10^4)$, even close to the upper critical dimension, $\rho_U = 4/3$. Indeed, the model has been used intensively in recent years for the study of the low temperature spin glass phase outside the mean-field theory \cite{leuzzi:09,leuzzi:11,larson:13,leuzzi:15}.

A very interesting, and still debated, question regards the existence of a spin glass phase transition in the non-mean-field region, $\rho_U<\rho<2$, when the external magnetic field is present. Standard methods of analysis, used up to now, provided non conclusive results, often interpreted in opposite ways \cite{larson:13}.

We have simulated the above model for two values of the long range exponent: $\rho=1.2$ which lies in  the mean-field region and $\rho=1.4$ which lies in the non-mean-field region. The zero-field critical temperatures are $T_c(h=0)=2.34(3)$ for $\rho=1.2$ and  $T_c(h=0)=1.970(2)$ for $\rho=1.4$ \cite{leuzzi:09}. We have simulated the equilibrium dynamics of systems of sizes $L\le 2^{13}$ with field values $h=0.1,0.2,0.3$ using the Parallel Tempering method~\cite{hukushima:96, marinari:98b}.
For each value of $\rho$ and $L$ we have simulated $O(10^5)$ different disordered samples.

\subsection{A new tool of analysis conditioning on the overlap}

Let $\boldsymbol{s}$ and $\boldsymbol{t}$ be the configurations of two real replicas evolved by Glauber dynamics, and suppose to have $M$ independent measurements, taken over many different samples. We define the conditional average of the overlap-overlap correlation function (or 4-point correlation function) as
\be
G(r|q) = \langle q_0 q_r|q\rangle \equiv
\frac{\sum_{i,j:|i-j|=r} \E[s_i t_i s_j t_j\,\delta_{q N,\mathbf{s}\cdot\mathbf{t}}]}{\sum_{i,j:|i-j|=r} \E[\delta_{q N,\mathbf{s}\cdot\mathbf{t}}]}\,,
\ee
where $\E[(\cdots)]$ stands for the empirical average over the $M$ measurements and $\boldsymbol{s}\cdot\boldsymbol{t}\equiv\sum_{k=1}^N s_k t_k$ (here the system size $N=L$ because we have a $d=1$ model).
$G(r|q)$ actually depends also on the system size $L$ and the temperature $T$, but we avoid making this explicit to keep notation lighter.
The standard correlation function is obtained taking the average over the conditioning variate
\be
G(r) = \int dq\,P(q)\,G(r|q)\,,
\ee
where $P(q) =\E[\delta_{qN,\mathbf{s}\cdot\mathbf{t}}]$ is the probability distribution of the overlap.

More precisely the correlation function, $G(r)$,  defined above is the total correlation function and can be expressed as a linear combination of the usual connected correlation functions defined in the replicon and longitudinal sectors: $G(r) = 2 G_\text{\tiny SG}(r) - G_\text{\tiny L}(r)$ \cite{dedominicis:06}.
In order to obtain the two connected correlations separately, we should simulate four replicas instead of two, and the analysis would become much more bothersome. However, this is not really needed since an eventual spin glass phase transition makes both $G(r)$ and $G_\text{\tiny SG}(r)$ decay critically.

The rationale beyond this conditional averaging is in the observation
\cite{parisi2012numerical,janus:14c} that measurements of a spin glass
model in field may have very large fluctuations even in the
paramagnetic phase. These huge fluctuations are associated with very
atypical overlap values: in the samples that one can simulate with
presently available computer resources the distribution of the
overlap, $P(q)$, has a tail in the region of small and even negative
overlaps; this tail will eventually disappear in the thermodynamic
limit, but is responsible for the huge finite size corrections.
Conditioning on the overlap, we expect $G(r|q)$ to have much weaker
fluctuations and so its estimate in the thermodynamic limit should be
less problematic.  Indeed we expect that, under the hypothesis of
replica equivalence \cite{parisi:00,marinari:00}, the conditional
correlation $G(r|q)$ should be self-averaging.  The main source of
fluctuations would remain in the $P(q)$, which is however very much
studied in the literature; moreover in the paramagnetic phase and in
the thermodynamic limit $P(q)=\delta(q-\qea)$, and we just need to
estimate $\qea$.

According to what we discussed in Section~\ref{sec:MF}, we expect that even in the paramagnetic phase the model may have a spin glass phase transition lowering the value of the conditioning overlap $q$. In order to detect this phase transition we can look for a critical overlap value, $q_c$, such that critical scaling holds in the conditioned susceptibility $\chi(q)$ at $q=q_c$.
To obtain the latter we Fourier transform the conditioned correlation function
\be
\widehat{G}(k|q) = \int dr\, e^{ikr} G(r|q)\;.
\ee
By definition $\widehat{G}(0|q)=q^2$ and so the best way to extract the large distance behavior of $G(r|q)$ is to look at $\widehat{G}(\kmin|q)$, with $\kmin=2\pi/L$. Thus  we define the susceptibility as
\be
\chi(q) \equiv \widehat{G}(\kmin|q)\;.
\ee

The model under study has power-law decaying interactions and this leads to the free propagator \cite{leuzzi:99}
\be
\widehat{G}(k)^{-1} \propto m^{2}+k^{\rho-1}\,.
\ee
The $\eta$ exponent is not renormalized in the non mean field region ($\rho_U<\rho<2$) and takes the value $\eta=3-\rho$, so  $2-\eta=\rho-1$. Therefore at criticality ($q=q_c$) we expect
\be
\label{eq:chic}
\chi(q_c)  \propto \left\{
\begin{array}{ll}
L^{\rho-1} &\text{  for  } \rho>\rho_U=4/3 \qquad \text{(non mean field)}\,,\\
L^{1/3} & \text{  for  } \rho\le\rho_U=4/3 \qquad \text{(mean field)}\,.
\end{array}
\right.
\ee
Being the exponent $\eta$ known, the value of $q_c$ can be obtained from the crossing of the curves $\chi(q) /L^{2-\eta}$ (or $\chi(q)/L^{1/3}$  in the mean field regime) plotted as a function of $q$. Notice that for $q<q_c$ we expect a power law divergence of $\chi(q)$ as a function of the lattice size, maybe with another power, and for $q>q_c$ the susceptibility $\chi(q)$ goes to a constant.

\subsection{Numerical results}

We present some results for $\rho=1.2$ (which belong to the mean field region) for comparison and the main results for $\rho=1.4$, which is in the interesting non-mean-field region, and far from the lower critical dimension.\footnote{The lower critical model has $\rho_L=2$ in absence of a field \cite{leuzzi:15}, but $\rho_L$ could decrease in a field \cite{leuzzi:13b}.}

\begin{figure}
\includegraphics[width=0.495\columnwidth]{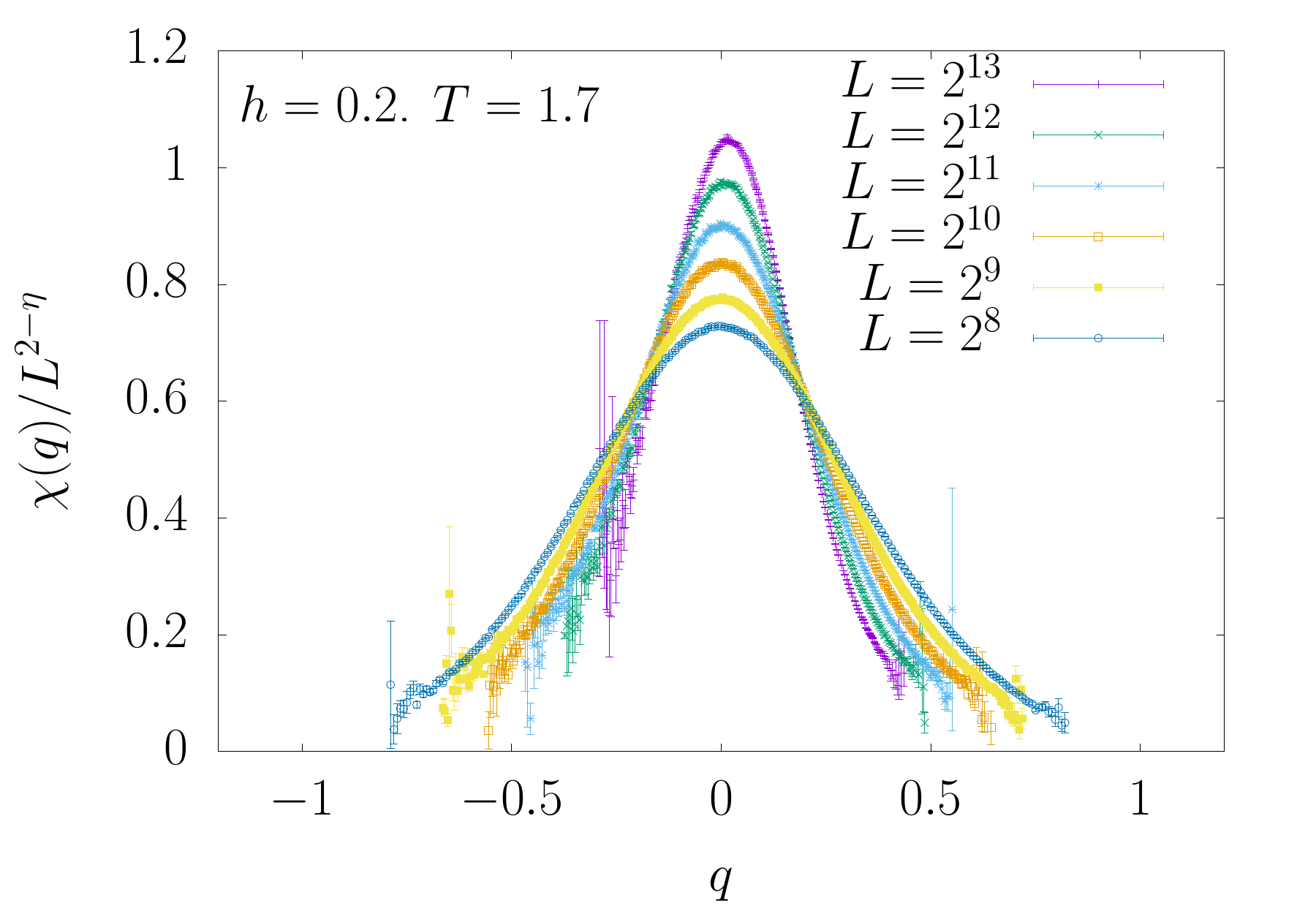}
\includegraphics[width=0.495\columnwidth]{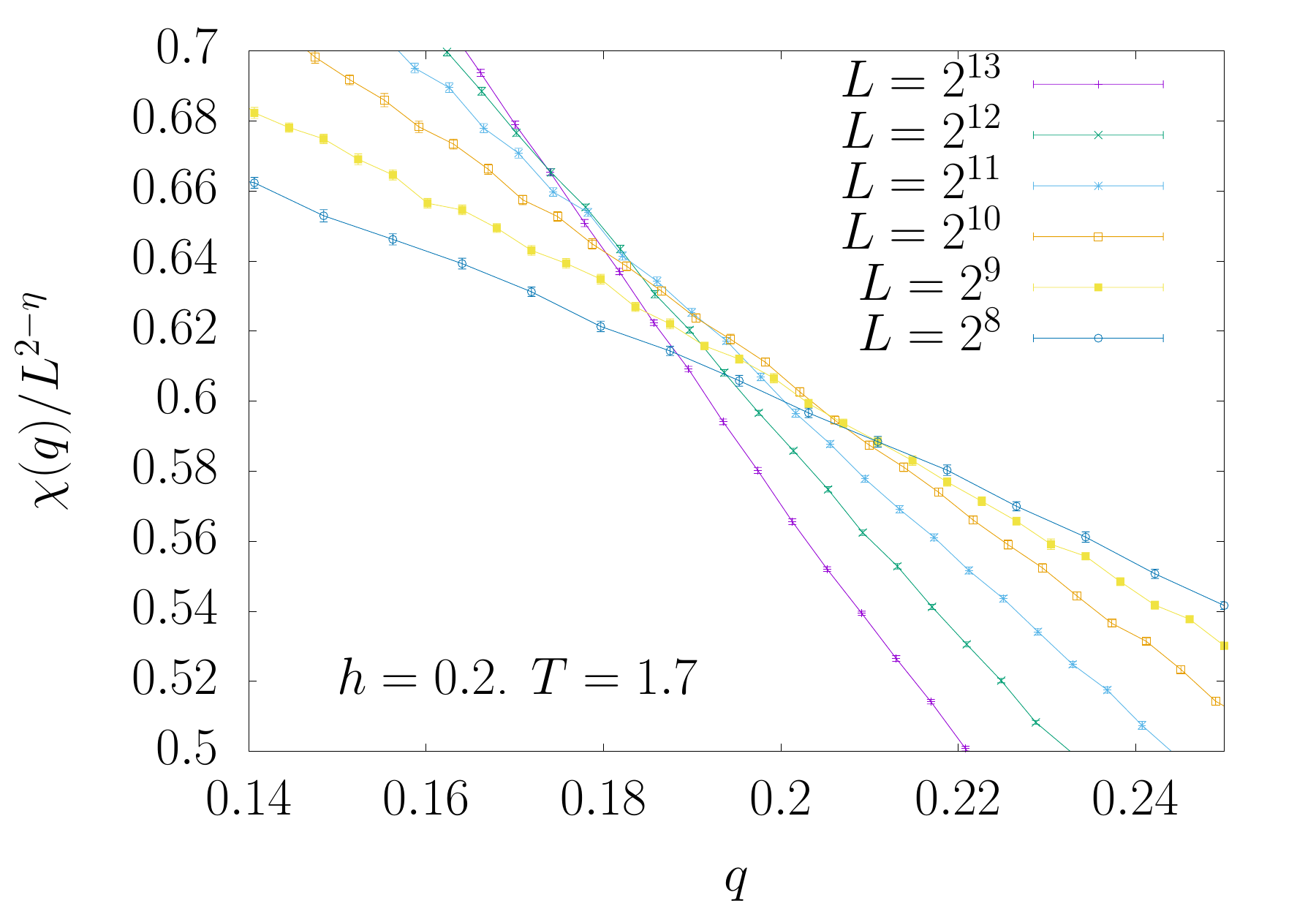}\\
\includegraphics[width=0.495\columnwidth]{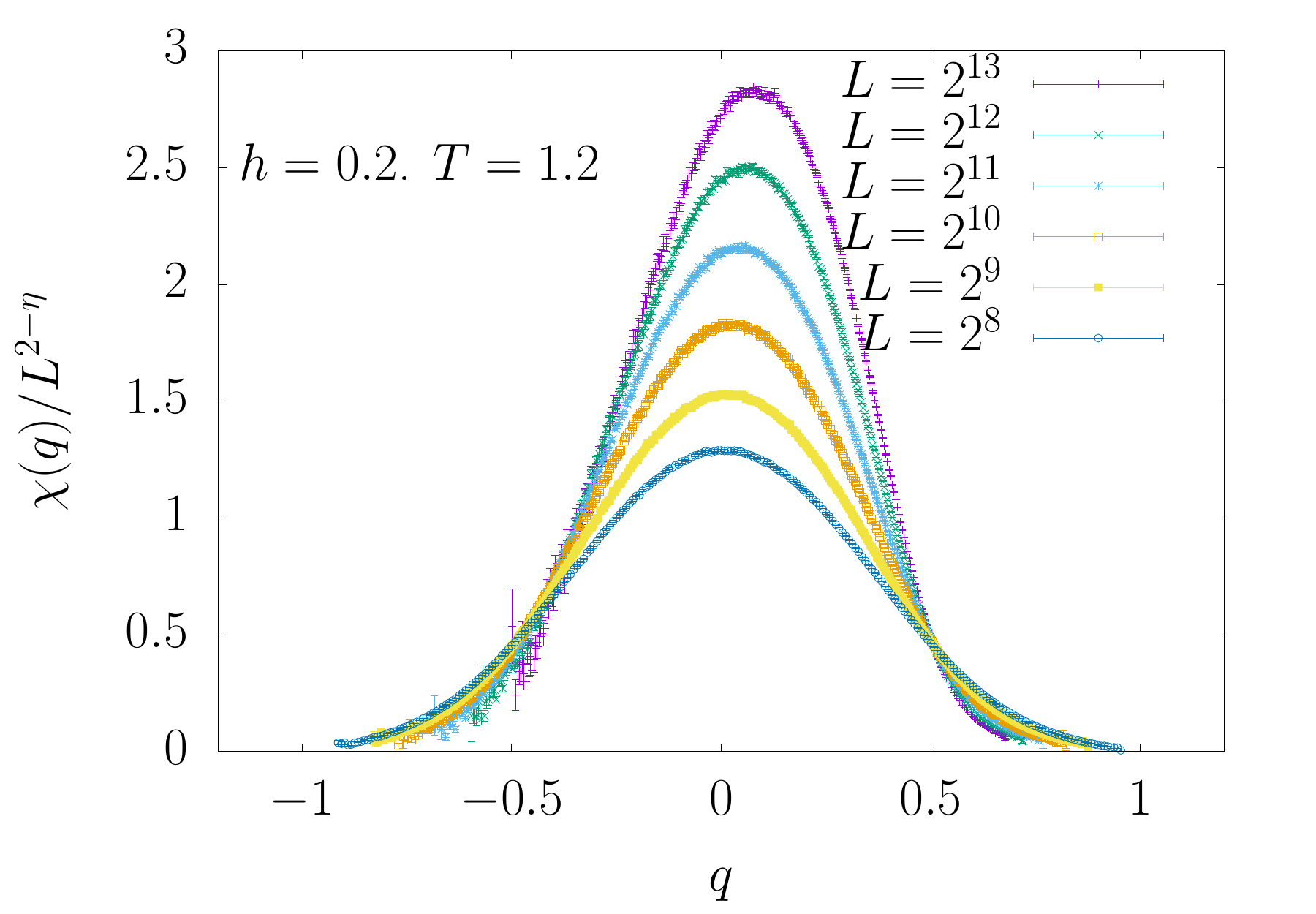}
\includegraphics[width=0.495\columnwidth]{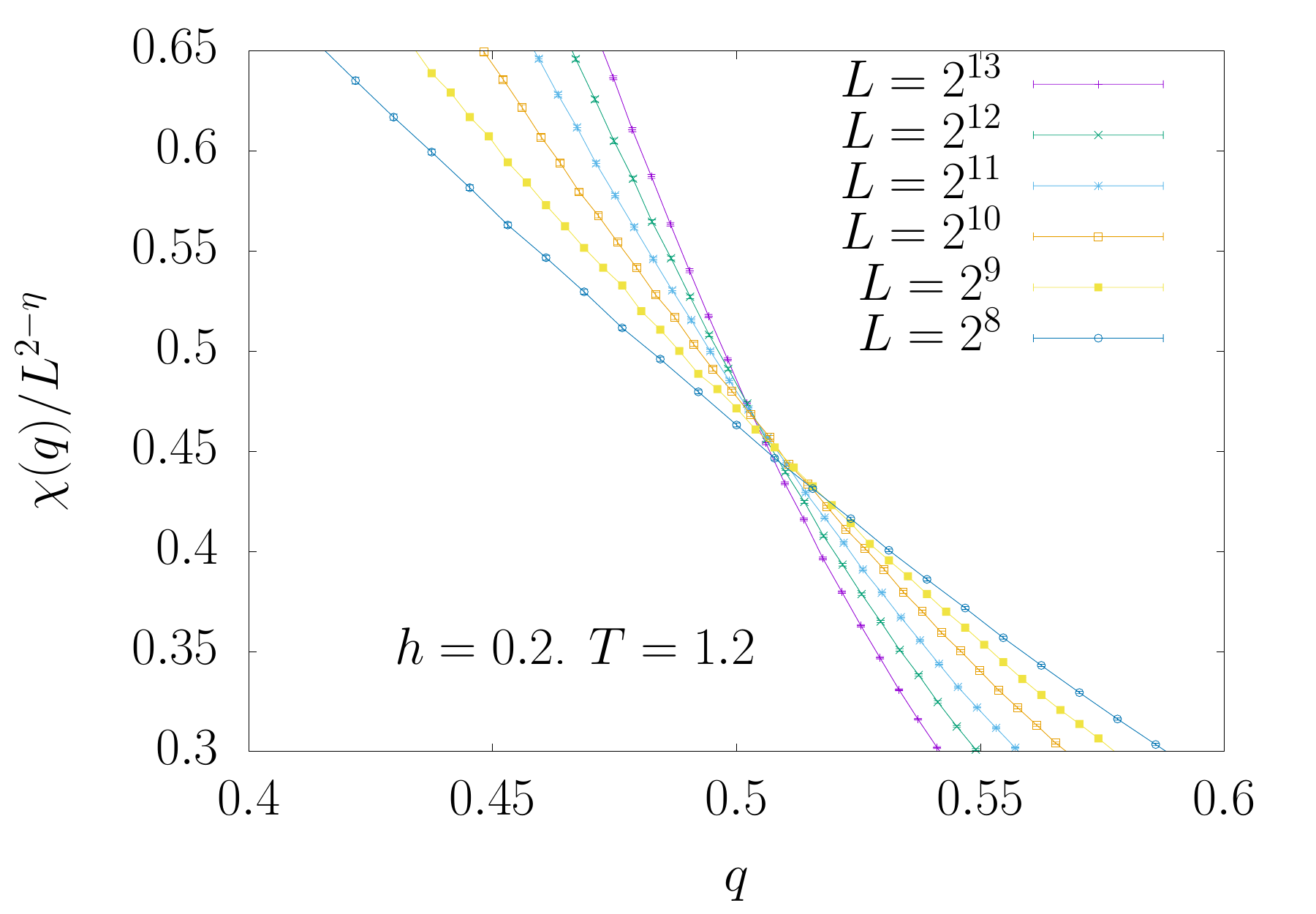}
\caption{$\chi(q)/L^{2-\eta}$ versus $q$ for $\rho=1.4$ (non mean field region) and six different lattice sizes. Data in the upper panels have been measured with $T=1.7$ and $h=0.2$, and belong to the paramagnetic phase~\cite{leuzzi:09}, thus showing that a transition to a spin glass phase can be induced by just decreasing the overlap between the replicas. In the bottom panels $T=1.2$ and $h=0.2$, which is near or inside the thermodynamic spin glass phase. The crossing point of the curves for different lattice sizes is always very neat as can be appreciated from the panels on the right that zoom on the crossing region.}
\label{fig:qc_estimate}
\end{figure}

In Figure~\ref{fig:qc_estimate} we plot the scaled susceptibilities,
$\chi(q)/L^{2-\eta}$, as a function of the conditioning overlap
$q$. The two plots are for $\rho=1.4$, $h=0.2$ and $T=1.7$ (top) and
$T=1.2$ (bottom). We remind that $T_c(h=0) \simeq 1.97$ for $\rho=1.4$
\cite{leuzzi:09}. All the curves, corresponding to sizes form $L=2^8$
to $L=2^{13}$, do intersect at well defined values of $q_c$, and
finite size corrections are not large and can be kept under control
with standard tools of analysis, making extrapolation to the
thermodynamic limit safe. It is worth noticing that the higher
temperature, $T=1.7$, is certainly in the paramagnetic phase,
according to previous estimate of $T_c=1.2(2)$ for $h=0.2$
\cite{leuzzi:09}. Nonetheless, lowering the overlap to atypical
values, smaller than the thermodynamically dominant value $\qea$,
clearly leads to a phase transition at $q_c$, in agreement with the
solution of the SK model.

We remind the reader that all the measurements have been taken during a standard Monte Carlo simulation, with no condition at all on the overlap among the two replicas. The conditioning on the overlap is imposed only during the off-line analysis.
The fact that curves in Fig.~\ref{fig:qc_estimate} have a good statistics in a broad $q$ range and become noisy only at the boundaries is a consequence of the very broad support of $P(q)$.
So, in some sense, the present analysis is exploiting in a positive way the huge fluctuations in the overlap that are present in spin glass models in a field.
One may complain that if the $P(q)$ would become narrow then the present analysis would fail. But in that case fluctuations are tiny and the standard analysis (without the conditioning) would provide a reliable answer.

\begin{figure} 
\includegraphics[width=0.495\columnwidth]{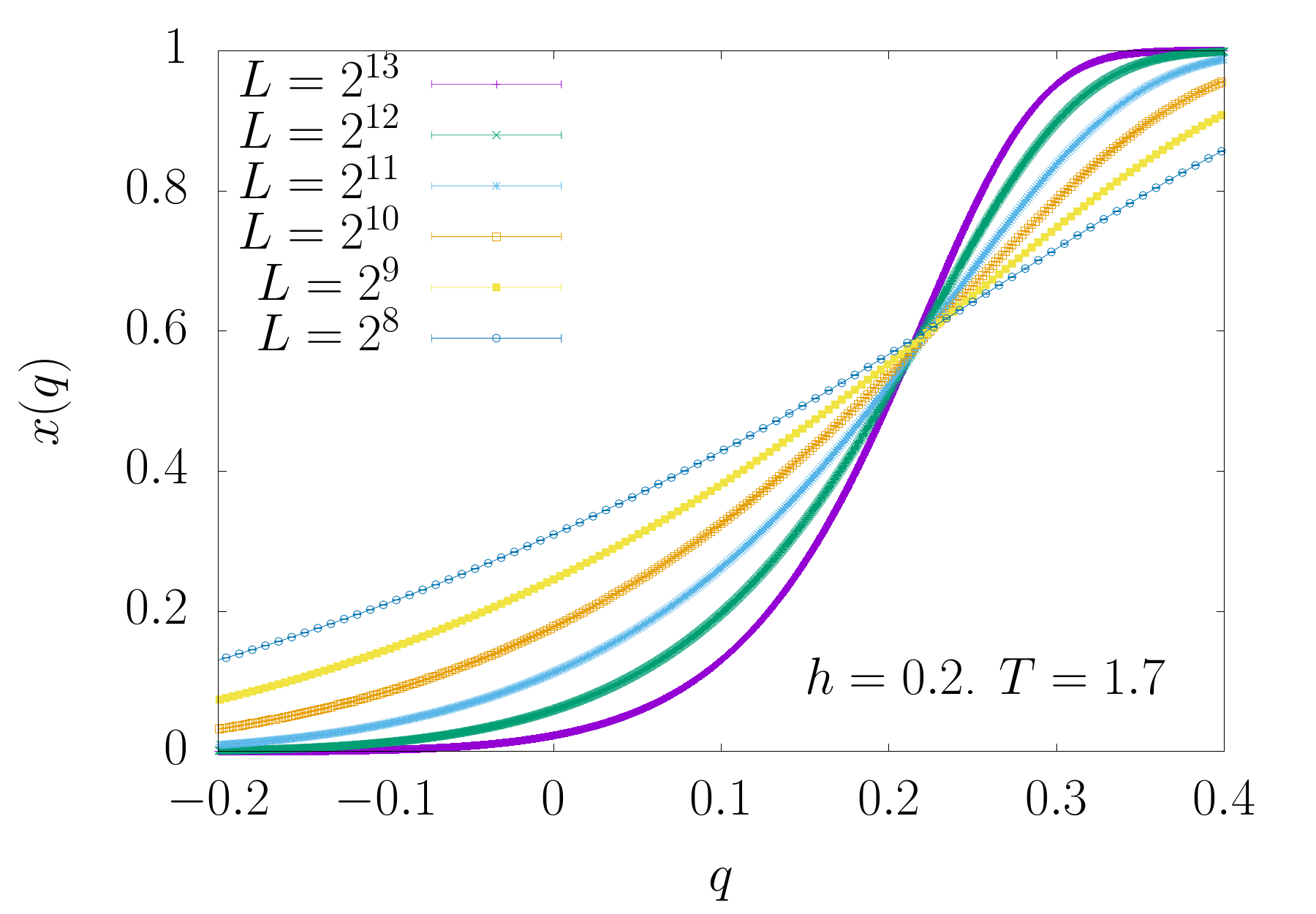}
\includegraphics[width=0.495\columnwidth]{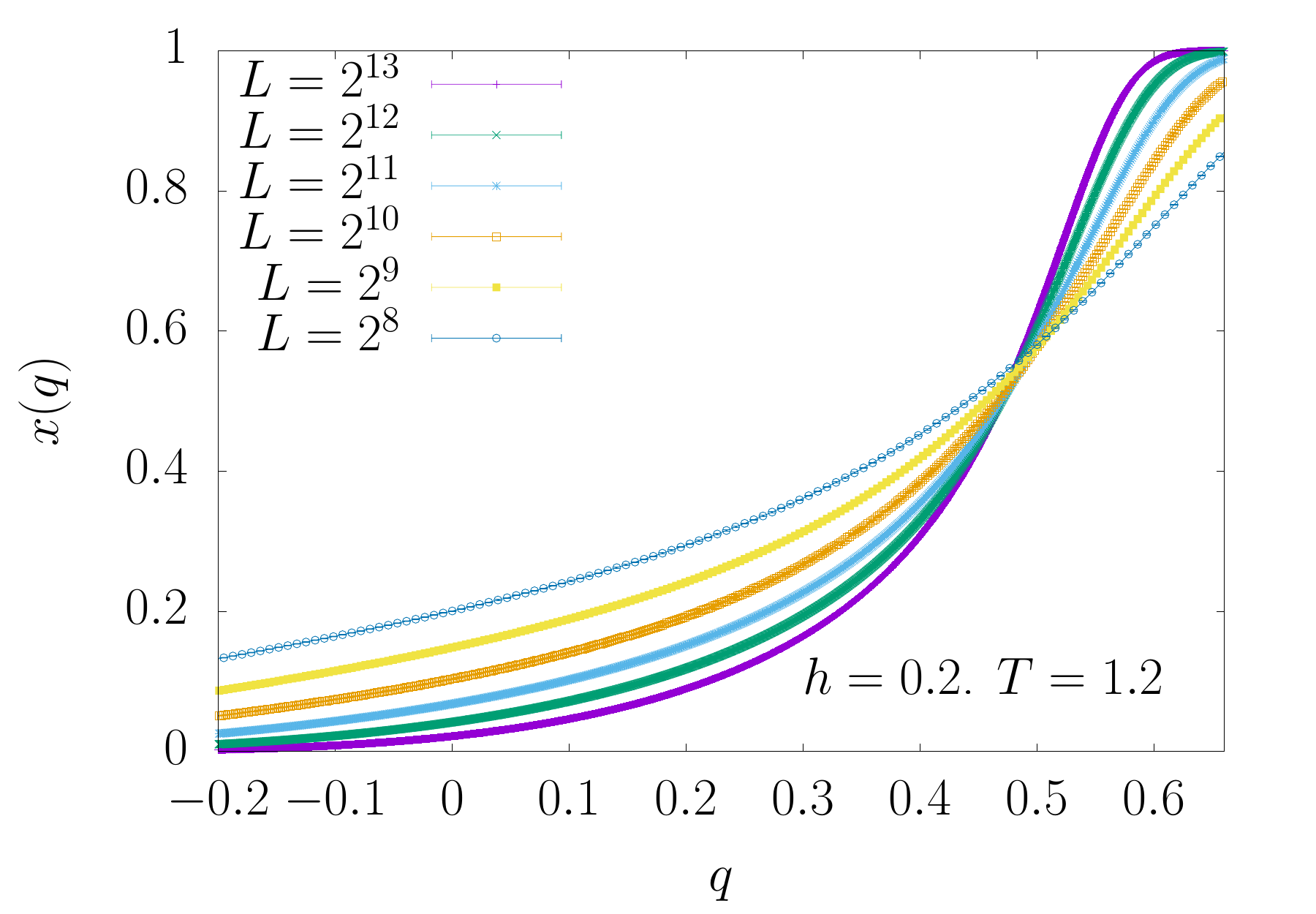}
\caption{The cumulative probability distribution $x(q)$ versus $q$ for $\rho=1.4$ (non mean field region), $h=0.2$ and two values of the temperature: $T=1.7$ (upper panel) and $T=1.2$ (lower panel). The estimate for $\qea$ comes from the crossing of these curves.}
\label{fig:xq}
\end{figure}

From the crossing point of the scaled susceptibilities shown in Fig.~\ref{fig:qc_estimate} we obtain reliable estimates for $q_c$. The estimate of $\qea$ can be obtained in two different ways as long as $T \ge T_c(h)$: 
\begin{itemize}
\item From the peak location in $P(q)$. 
\item From the crossing points of the cumulative functions $x(q) \equiv \int_{-1}^q dq'\,P(q')$.
\end{itemize}
We have found that the second method presents much weaker finite size corrections, and thus we have used it to estimate $\qea$ (see Fig.~\ref{fig:xq}).
It is worth noticing that this second method, although much more accurate as long as $T \ge T_c(h)$, does not return the correct result for $T<T_c(h)$.
Indeed, in the latter case the median value estimated from the crossing of the cumulative functions $x(q)$ is slightly lower than the location of the peak in the $P(q)$.
Nonetheless, for the purpose of locating an eventual critical temperature $T_c(h)$ a reliable estimate of $\qea$ in the region $T \ge T_c(h)$ is enough.

\begin{figure}
\begin{tabular}{c}
\includegraphics[width=0.489\columnwidth]{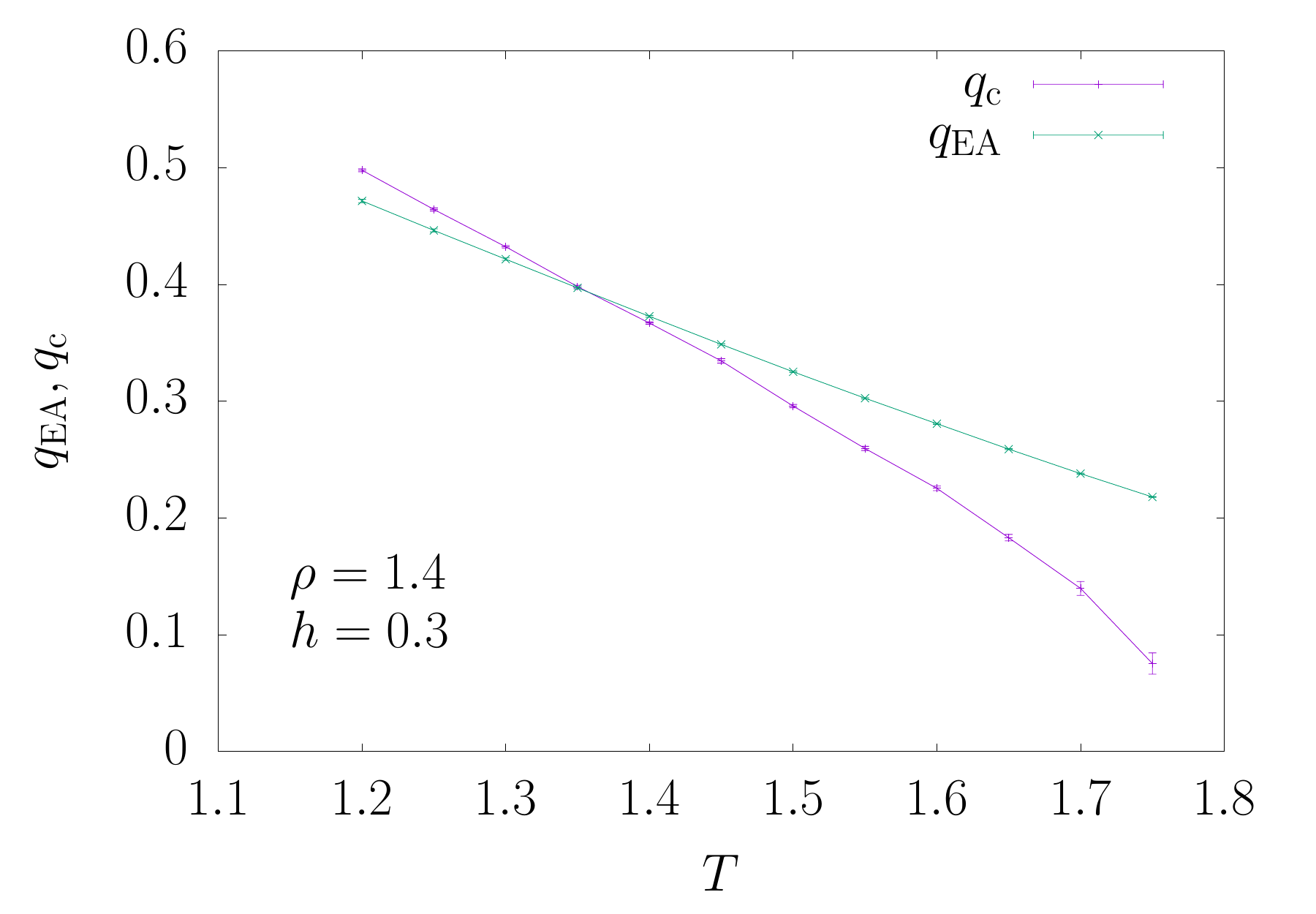}\\
\includegraphics[width=0.489\columnwidth]{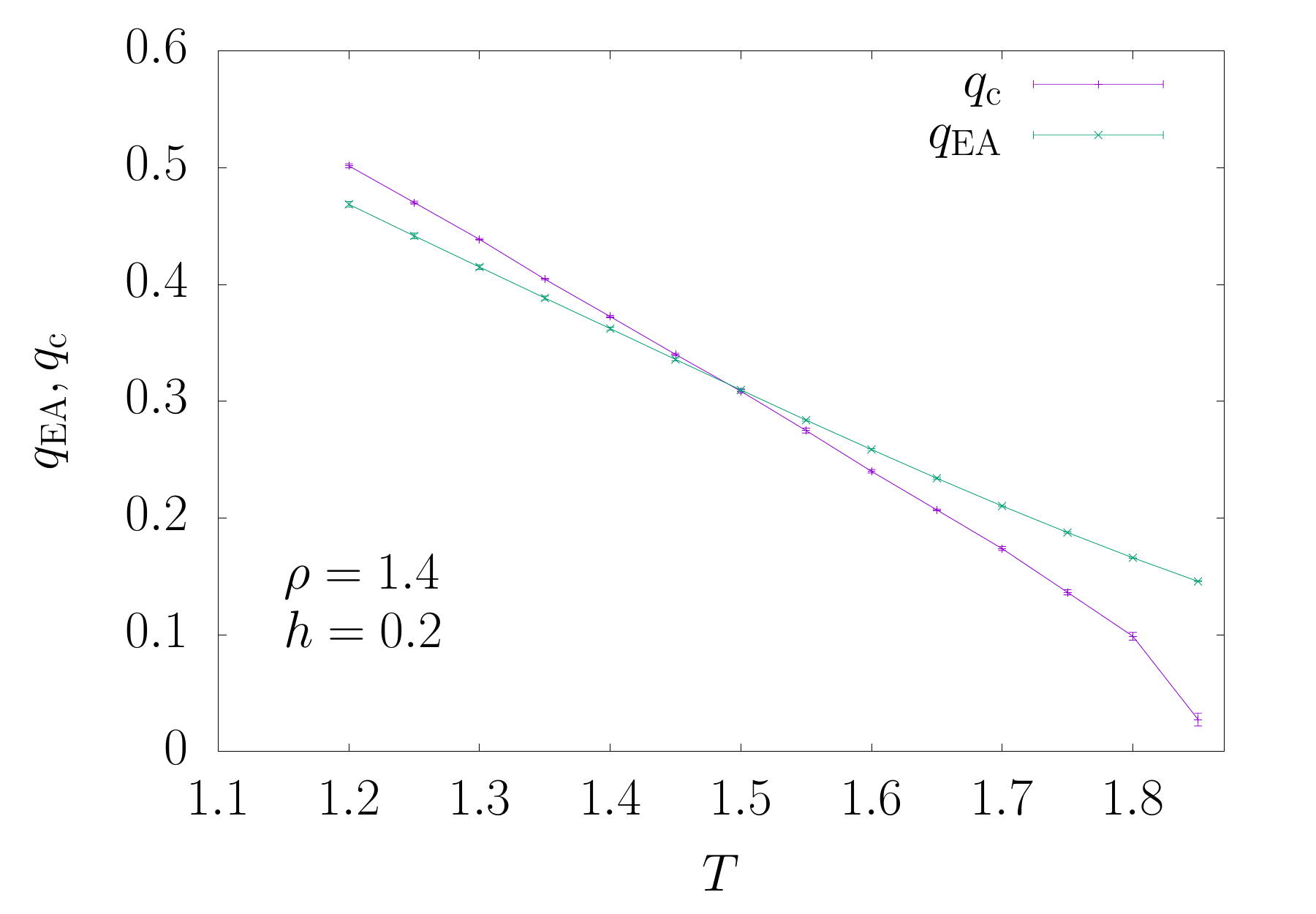}\\
\includegraphics[width=0.489\columnwidth]{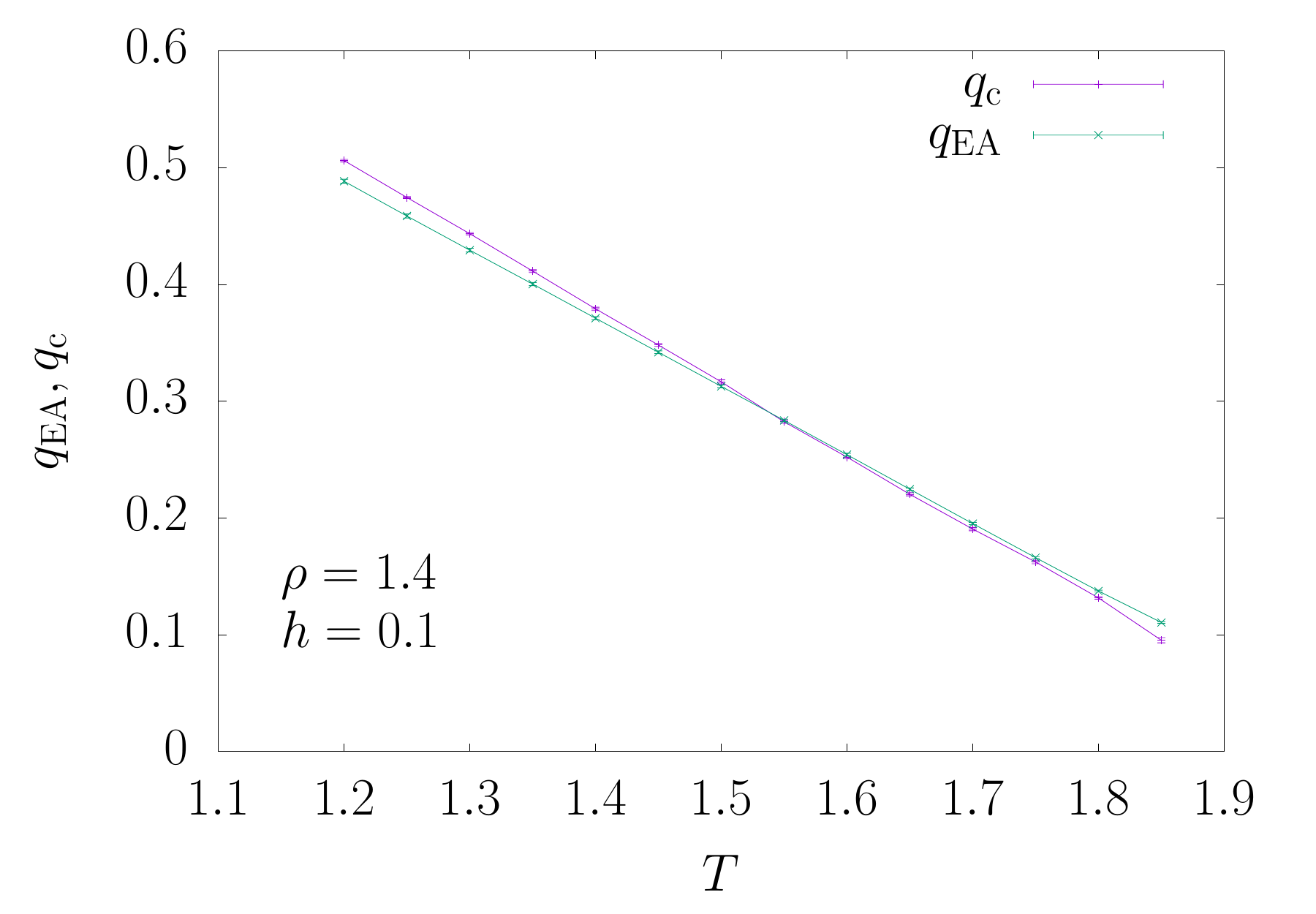}
\end{tabular}
\begin{tabular}{c}
\includegraphics[width=0.489\columnwidth]{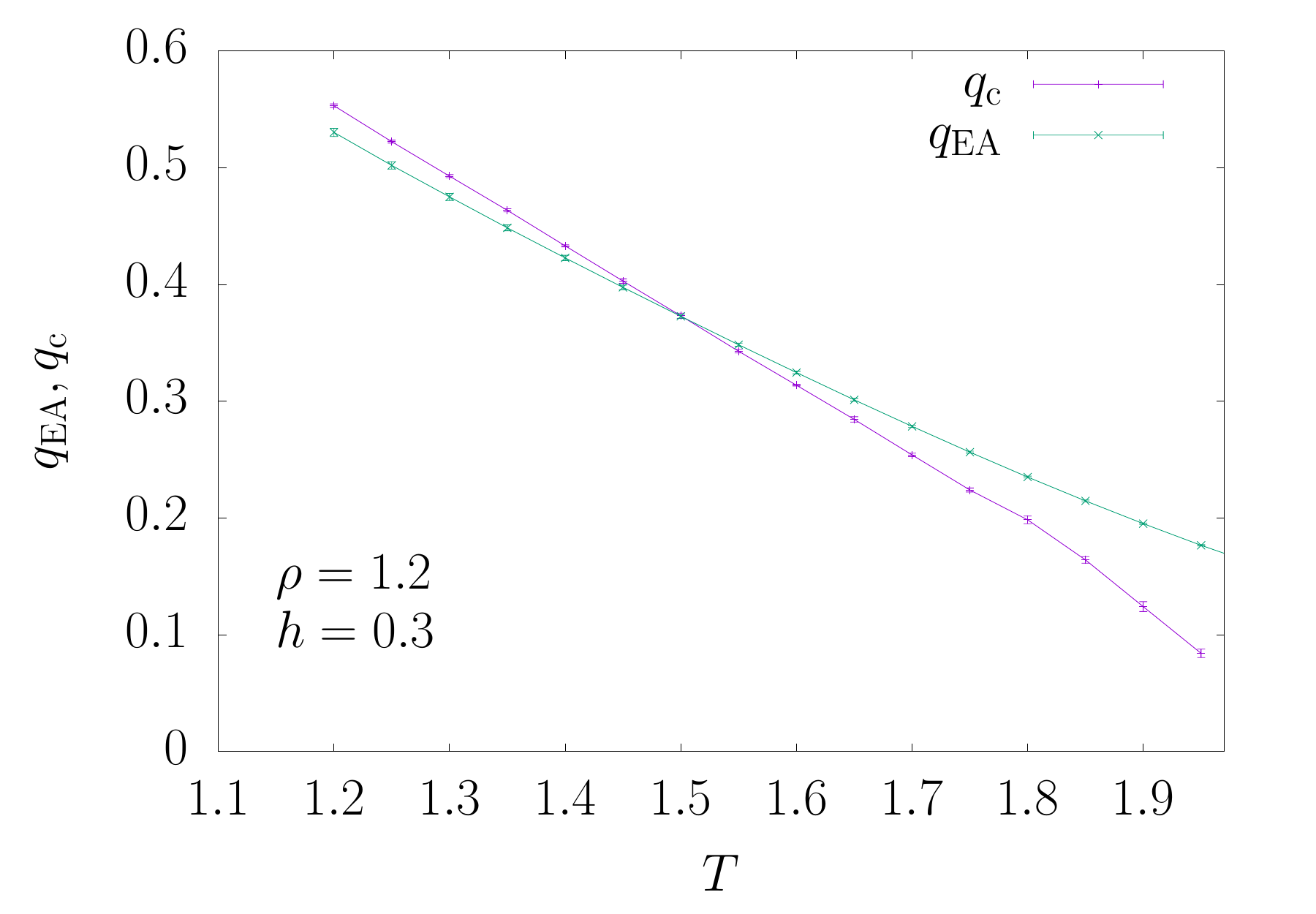}\\
\includegraphics[width=0.489\columnwidth]{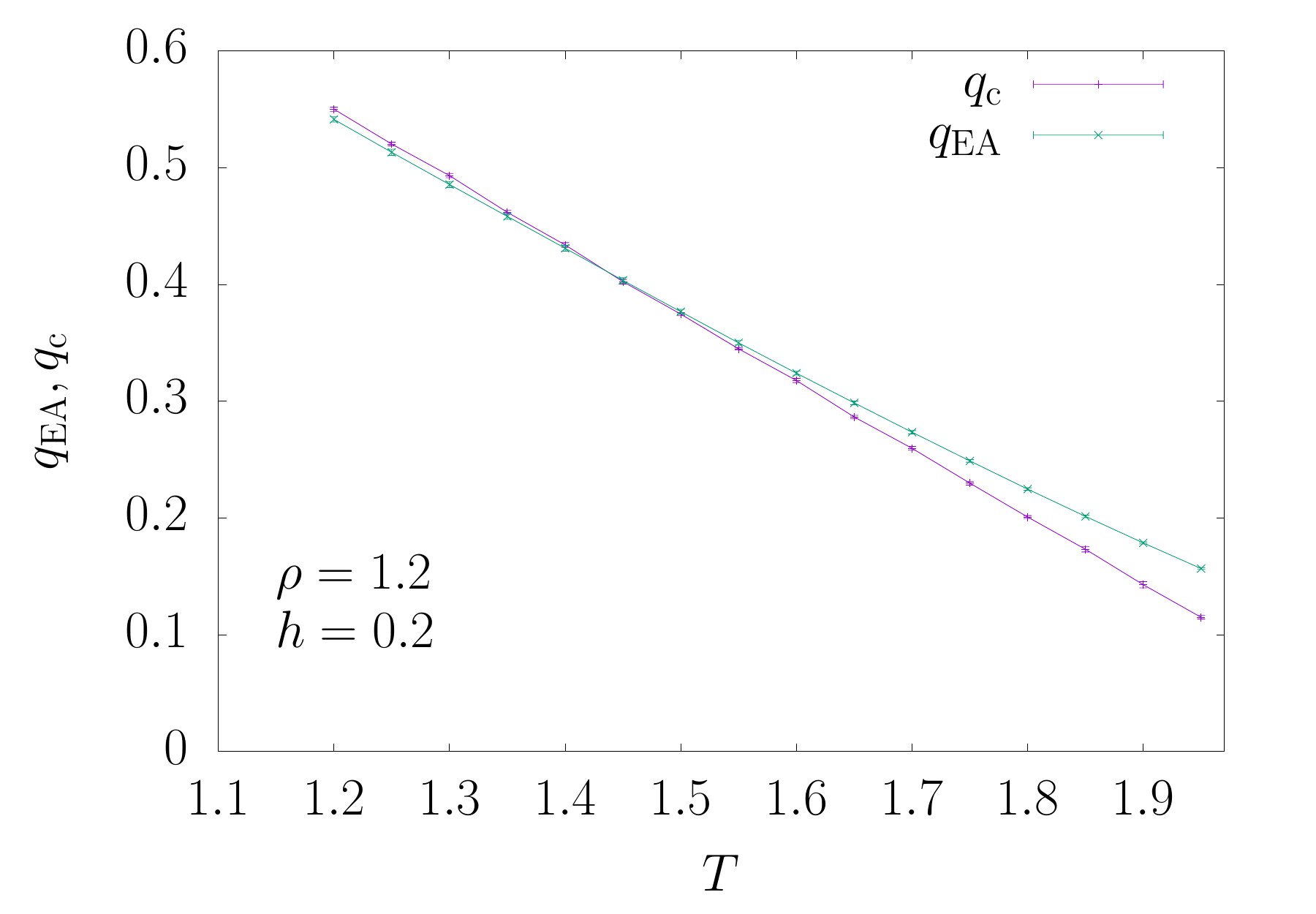}
\end{tabular}
\caption{Behavior of $q_c(T)$ and $\qea(T)$ for $\rho=1.4$ (left panels, non mean field regime) and $\rho=1.2$ (right panels, mean field regime), and several values of the magnetic field. The crossing (or merging) of the curves identifies the thermodynamic phase transition to the spin glass phase (dAT line) because $q_c<\qea$ holds in the paramagnetic phase. Data shown are for the largest sizes ($L=2^{12}$ and $L=2^{13}$).}
\label{fig:qea_qc}
\end{figure}

We show in Fig.~\ref{fig:qea_qc} the estimates of $q_c$ and $\qea$
obtained for $\rho=1.4$ (left panels) and $\rho=1.2$ (right panels)
with different values of the magnetic field and the largest sizes
($L=2^{12}$ and $L=2^{13}$).  As long as $\qea>q_c$ in the
thermodynamic limit, the model is not critical and belongs to the
paramagnetic phase.  Exactly when $\qea=q_c$ the model satisfies
critical scaling in the thermodynamic limit, that is, it belongs to
the critical dAT line $T_c(h)$.  For $\qea<q_c$ we are below $T_c(h)$
and thus the estimate of $\qea$ is no longer correct (most probably
the correct estimate of $\qea$ is $q_c$ itself).

With the aim of estimating the critical dAT line $T_c(h)$ one should compute the crossing points between the $\qea(T)$ and $q_c(T)$ curves shown in Fig.~\ref{fig:qea_qc}.
For fields large enough (e.g.\ $h\ge 0.2$ for $\rho=1.4$) the crossing of the $\qea(T)$ and $q_c(T)$ curves is very clear, and thus the identification of the critical dAT temperature $T_c(h)$ is highly reliable.
On the contrary, for very small fields (e.g.\ $h=0.1$ for $\rho=1.4$) the $\qea(T)$ and $q_c(T)$ curves tend to merge, rather than crossing each other and thus the estimate of $T_c(h)$ is noisy. This comes with no surprises given that we expect the two curves to coincide in the $h=0$ limit.
In this case we adopt the rule of estimating the critical temperature as the highest temperature at which the difference is compatible with zero within one standard deviation (and this may provide a bias towards too large values).

We observe also that the analysis works much better for $\rho=1.4$ than for $\rho=1.2$, and this is a good news, given that we are mainly interested in locating the dAT line in the non mean field regime; the data for $\rho=1.2$ have been shown mainly for comparison purposes, but the actual estimates of the critical point in this case are very noisy as was already noticed in Ref.~\cite{leuzzi:09}.

\begin{table}[h]
  \centering
\begin{tabular}{c l l l}
\firsthline
\hline
\multicolumn{4}{c}{$\rho=1.4$}\\
\cline{1-4}
\multicolumn{1}{c}{}&\multicolumn{1}{c}{$h=0.1$} &\multicolumn{1}{c}{$h=0.2$} &\multicolumn{1}{c}{$h=0.3$} \\
\cline{2-4} 
\multicolumn{1}{c}{$\log_2 L$}&\multicolumn{1}{c}{$T_c$} &\multicolumn{1}{c}{$T_c$} &\multicolumn{1}{c}{$T_c$} \\
\hline
8        & 1.88(1)     & 1.56(6) & 1.31(4)    \\
9        & 1.89(3)     & 1.44(6) & 1.39(3)     \\
10       & 1.85(1)     & 1.47(2) & 1.40(1)    \\
11       & 1.40(3)     & 1.53(1) & 1.39(3)      \\
12       & 1.57(9)     & 1.51(1) & 1.37(1)     \\
\hline
FSSA   &  1.67(7)      & 1.2(2)  &              \\
\hline
\lasthline
\end{tabular}
\hspace{1cm}
\begin{tabular}{c l l}
\firsthline
\hline
\multicolumn{3}{c}{$\rho=1.2$}\\
\cline{1-3}
\multicolumn{1}{c}{}&\multicolumn{1}{c}{$h=0.2$} &\multicolumn{1}{c}{$h=0.3$} \\
\cline{2-3} 
\multicolumn{1}{c}{$\log_2 L$}&\multicolumn{1}{c}{$T_c$} &\multicolumn{1}{c}{$T_c$} \\\hline
8        &             & 1.47(10)      \\
9        & 1.36(5)     & 1.38(5)      \\
10       & 1.4(1)      & 1.43(4)      \\
11       & 1.48(5)     & 1.47(3)       \\
12       & 1.51(5)     & 1.53(2)       \\
\hline
FSSA      & 1.4(2)     & 1.5(4) \\
\hline
\lasthline
\end{tabular}
\caption{Values of the critical temperature obtained from the crossing points of the curves $\qea(T)$ and $q_c(T)$, which have been computed using data from lattices $L$ and $2L$. Left table is for $\rho=1.4$ (non-mean-field regime) and right table is for $\rho=1.2$ (mean-field regime). In the row labelled FSSA we have reported the critical temperatures obtained in Ref.~\cite{leuzzi:09} using Finite Size Scaling Analysis.
  \label{table:tc}}
\end{table}  

In order to highlight how stable is the analysis and to check for finite size corrections we have computed the curves $\qea(T)$ and $q_c(T)$ from the data of just two sizes, namely $L$ and $2L$. In practice, $\qea$ is obtained from $x_L(\qea)=x_{2L}(\qea)$ while $q_c$ satisfies $\chi_L(q_c)/L^{2-\eta}=\chi_{2L}(q_c)/(2L)^{2-\eta}$. The \emph{finite size estimate} of the critical temperature corresponds to the crossing of these $\qea(T)$ and $q_c(T)$ obtained from sizes $L$ and $2L$.
We report these estimates in Table~\ref{table:tc} and we notice that finite size corrections are smooth and under control, but for the smallest fields (but this is expected given the observation that at small fields estimates of $T_c(h)$ are much more noisy).

\section{Conclusions}

In this work we have studied, both analytically and numerically, the phase transition from the paramagnetic to the spin glass phase that takes place in a spin glass model when two real replicas are forced to have an overlap smaller than the equilibrium one. 

On the one hand we have solved the SK model close to its critical point (under the so-called truncated approximation) and we have shown analytically that in the paramagnetic phase in a field ($h>0$ and $T_c(h)<T<T_c(0)$) lowering the overlap $p_d$ between two real replicas leads to a phase transition at $p_d^*$ where the symmetry between replicas gets spontaneously broken.
In the whole paramagnetic phase the inequality $p_d^*<\qea$ holds, while $p_d^*=\qea$ identifies the critical dAT line.

On the other hand, we have performed an analysis of data from Monte Carlo simulations of a proxy of a finite-dimensional spin glass model conditioning on the value of the overlap between the two simulated replicas. We have cleanly estimated a value $q_c$ of the overlap where the model undergoes a phase transition and it is natural to identify the numerically estimated $q_c$ with the analytically predicted $p_d^*$. The comparison of $q_c$ with the typical overlap $\qea$ turned out to be a very reliable way to estimate the critical dAT line $T_c(h)$.

In conclusion we have provided a solid evidence for the existence of a critical dAT line in a finite-dimensional spin glass model in a field by using a new tool of analysis inspired by the presence of a phase transition in the overlap between two real replicas.

\section{Acknowledgements}

We thank E. Marinari and V. Martin-Mayor for interesting discussions.

This work was supported by the European Research Council under the
European Unions Horizon 2020 research and innovation programme (grant
No. 694925, G. Parisi), by the project “Meccanica statistica e
complessità”, a research grant funded by PRIN 2015 (Agreement
no. 2015K7KK8L), by Ministerio de Econom\'{\i}a y Competitividad,
Agencia Estatal de Investigaci\'on, and Fondo Europeo de Desarrollo
Regional (FEDER) (Spain and European Union) through grants
No.\ FIS2016-76359-P and by Junta de Extremadura (Spain) through
grants No.\ GRU18079 and No.\ IB16013 (partially funded by FEDER).


\begin{thebibliography}{36}
\expandafter\ifx\csname natexlab\endcsname\relax\def\natexlab#1{#1}\fi
\expandafter\ifx\csname bibnamefont\endcsname\relax
  \def\bibnamefont#1{#1}\fi
\expandafter\ifx\csname bibfnamefont\endcsname\relax
  \def\bibfnamefont#1{#1}\fi
\expandafter\ifx\csname citenamefont\endcsname\relax
  \def\citenamefont#1{#1}\fi
\expandafter\ifx\csname url\endcsname\relax
  \def\url#1{\texttt{#1}}\fi
\expandafter\ifx\csname urlprefix\endcsname\relax\def\urlprefix{URL }\fi
\providecommand{\bibinfo}[2]{#2}
\providecommand{\eprint}[2][]{\url{#2}}

\bibitem[{\citenamefont{de~Almeida and Thouless}(1978)}]{dealmeida:78}
\bibinfo{author}{\bibfnamefont{J.~R.~L.} \bibnamefont{de~Almeida}}
  \bibnamefont{and} \bibinfo{author}{\bibfnamefont{D.~J.}
  \bibnamefont{Thouless}}, \bibinfo{journal}{J. Phys. A: Math. Gen.}
  \textbf{\bibinfo{volume}{11}}, \bibinfo{pages}{983} (\bibinfo{year}{1978}),
  \urlprefix\url{http://stacks.iop.org/0305-4470/11/i=5/a=028}.

\bibitem[{\citenamefont{{Caracciolo, Sergio}
  et~al.}(1990)\citenamefont{{S. Caracciolo}, {G. Parisi},
  {Patarnello, Stefano}, and {Sourlas Nicolas}}}]{caracciolo:90}
\bibinfo{author}{\bibnamefont{{S. Caracciolo}}},
  \bibinfo{author}{\bibnamefont{{G. Parisi}}},
  \bibinfo{author}{\bibnamefont{{S. Patarnello}}}, \bibnamefont{and}
  \bibinfo{author}{\bibnamefont{{N. Sourlas}}}, \bibinfo{journal}{J.
  Phys. France} \textbf{\bibinfo{volume}{51}}, \bibinfo{pages}{1877}
  (\bibinfo{year}{1990}),
  \urlprefix\url{https://doi.org/10.1051/jphys:0199000510170187700}.

\bibitem[{\citenamefont{{David A. Huse} and {Daniel S.
  Fisher}}(1991)}]{huse:91}
\bibinfo{author}{\bibnamefont{{D. A. Huse}}} \bibnamefont{and}
  \bibinfo{author}{\bibnamefont{{D. S. Fisher}}}, \bibinfo{journal}{J.
  Phys. I France} \textbf{\bibinfo{volume}{1}}, \bibinfo{pages}{621}
  (\bibinfo{year}{1991}), \urlprefix\url{https://doi.org/10.1051/jp1:1991157}.

\bibitem[{\citenamefont{{Sergio Caracciolo} et~al.}(1991)\citenamefont{{Sergio
  Caracciolo}, {Giorgio Parisi}, {Stefano Patarnello}, and {Nicolas
  Sourlas}}}]{caracciolo:91}
\bibinfo{author}{\bibnamefont{{S. Caracciolo}}},
  \bibinfo{author}{\bibnamefont{{G. Parisi}}},
  \bibinfo{author}{\bibnamefont{{S. Patarnello}}}, \bibnamefont{and}
  \bibinfo{author}{\bibnamefont{{N. Sourlas}}}, \bibinfo{journal}{J. Phys.
  I France} \textbf{\bibinfo{volume}{1}}, \bibinfo{pages}{627}
  (\bibinfo{year}{1991}), \urlprefix\url{https://doi.org/10.1051/jp1:1991158}.

\bibitem[{\citenamefont{{J.C. Ciria} et~al.}(1993)\citenamefont{{J.C. Ciria},
  {G. Parisi}, {F. Ritort}, and {J.J. Ruiz-Lorenzo}}}]{ciria:93}
\bibinfo{author}{\bibnamefont{{J.C. Ciria}}}, \bibinfo{author}{\bibnamefont{{G.
  Parisi}}}, \bibinfo{author}{\bibnamefont{{F. Ritort}}}, \bibnamefont{and}
  \bibinfo{author}{\bibnamefont{{J.J. Ruiz-Lorenzo}}}, \bibinfo{journal}{J.
  Phys. I France} \textbf{\bibinfo{volume}{3}}, \bibinfo{pages}{2207}
  (\bibinfo{year}{1993}), \urlprefix\url{https://doi.org/10.1051/jp1:1993241}.

\bibitem[{\citenamefont{Parisi et~al.}(1998)\citenamefont{Parisi,
  Ricci-Tersenghi, and Ruiz-Lorenzo}}]{parisi:98b}
\bibinfo{author}{\bibfnamefont{G.}~\bibnamefont{Parisi}},
  \bibinfo{author}{\bibfnamefont{F.}~\bibnamefont{Ricci-Tersenghi}},
  \bibnamefont{and} \bibinfo{author}{\bibfnamefont{J.~J.}
  \bibnamefont{Ruiz-Lorenzo}}, \bibinfo{journal}{Phys. Rev. B}
  \textbf{\bibinfo{volume}{57}}, \bibinfo{pages}{13617} (\bibinfo{year}{1998}),
  \eprint{arXiv:cond-mat/9711122}.

\bibitem[{\citenamefont{Marinari
  et~al.}(1998{\natexlab{a}})\citenamefont{Marinari, Parisi, and
  Zuliani}}]{marinari:98e}
\bibinfo{author}{\bibfnamefont{E.}~\bibnamefont{Marinari}},
  \bibinfo{author}{\bibfnamefont{G.}~\bibnamefont{Parisi}}, \bibnamefont{and}
  \bibinfo{author}{\bibfnamefont{F.}~\bibnamefont{Zuliani}},
  \bibinfo{journal}{J. Phys. A: Math. and Gen.} \textbf{\bibinfo{volume}{31}},
  \bibinfo{pages}{1181} (\bibinfo{year}{1998}{\natexlab{a}}),
  \eprint{arXiv:cond-mat/9703253}.

\bibitem[{\citenamefont{Marinari
  et~al.}(1998{\natexlab{b}})\citenamefont{Marinari, Naitza, and
  Parisi}}]{marinari:98g}
\bibinfo{author}{\bibfnamefont{E.}~\bibnamefont{Marinari}},
  \bibinfo{author}{\bibfnamefont{C.}~\bibnamefont{Naitza}}, \bibnamefont{and}
  \bibinfo{author}{\bibfnamefont{G.}~\bibnamefont{Parisi}},
  \bibinfo{journal}{Journal of Physics A: Mathematical and General}
  \textbf{\bibinfo{volume}{31}}, \bibinfo{pages}{6355}
  (\bibinfo{year}{1998}{\natexlab{b}}).

\bibitem[{\citenamefont{Marinari
  et~al.}(1998{\natexlab{c}})\citenamefont{Marinari, Naitza, Zuliani, Parisi,
  Picco, and Ritort}}]{marinari:98h}
\bibinfo{author}{\bibfnamefont{E.}~\bibnamefont{Marinari}},
  \bibinfo{author}{\bibfnamefont{C.}~\bibnamefont{Naitza}},
  \bibinfo{author}{\bibfnamefont{F.}~\bibnamefont{Zuliani}},
  \bibinfo{author}{\bibfnamefont{G.}~\bibnamefont{Parisi}},
  \bibinfo{author}{\bibfnamefont{M.}~\bibnamefont{Picco}}, \bibnamefont{and}
  \bibinfo{author}{\bibfnamefont{F.}~\bibnamefont{Ritort}},
  \bibinfo{journal}{Phys. Rev. Lett.} \textbf{\bibinfo{volume}{81}},
  \bibinfo{pages}{1698} (\bibinfo{year}{1998}{\natexlab{c}}),
  \urlprefix\url{https://link.aps.org/doi/10.1103/PhysRevLett.81.1698}.

\bibitem[{\citenamefont{Houdayer and Martin}(1999)}]{houdayer:99}
\bibinfo{author}{\bibfnamefont{J.}~\bibnamefont{Houdayer}} \bibnamefont{and}
  \bibinfo{author}{\bibfnamefont{O.~C.} \bibnamefont{Martin}},
  \bibinfo{journal}{Phys. Rev. Lett.} \textbf{\bibinfo{volume}{82}},
  \bibinfo{pages}{4934} (\bibinfo{year}{1999}),
  \urlprefix\url{https://link.aps.org/doi/10.1103/PhysRevLett.82.4934}.

\bibitem[{\citenamefont{Marinari
  et~al.}(2000{\natexlab{a}})\citenamefont{Marinari, Parisi, and
  Zuliani}}]{marinari:00d}
\bibinfo{author}{\bibfnamefont{E.}~\bibnamefont{Marinari}},
  \bibinfo{author}{\bibfnamefont{G.}~\bibnamefont{Parisi}}, \bibnamefont{and}
  \bibinfo{author}{\bibfnamefont{F.}~\bibnamefont{Zuliani}},
  \bibinfo{journal}{Phys. Rev. Lett.} \textbf{\bibinfo{volume}{84}},
  \bibinfo{pages}{1056} (\bibinfo{year}{2000}{\natexlab{a}}),
  \urlprefix\url{https://link.aps.org/doi/10.1103/PhysRevLett.84.1056}.

\bibitem[{\citenamefont{Houdayer and Martin}(2000)}]{houdayer:00}
\bibinfo{author}{\bibfnamefont{J.}~\bibnamefont{Houdayer}} \bibnamefont{and}
  \bibinfo{author}{\bibfnamefont{O.~C.} \bibnamefont{Martin}},
  \bibinfo{journal}{Phys. Rev. Lett.} \textbf{\bibinfo{volume}{84}},
  \bibinfo{pages}{1057} (\bibinfo{year}{2000}),
  \urlprefix\url{https://link.aps.org/doi/10.1103/PhysRevLett.84.1057}.

\bibitem[{\citenamefont{Cruz et~al.}(2003)\citenamefont{Cruz, Fern\'andez,
  Jim\'enez, Ruiz-Lorenzo, and Taranc\'on}}]{cruz:03}
\bibinfo{author}{\bibfnamefont{A.}~\bibnamefont{Cruz}},
  \bibinfo{author}{\bibfnamefont{L.~A.} \bibnamefont{Fern\'andez}},
  \bibinfo{author}{\bibfnamefont{S.}~\bibnamefont{Jim\'enez}},
  \bibinfo{author}{\bibfnamefont{J.~J.} \bibnamefont{Ruiz-Lorenzo}},
  \bibnamefont{and}
  \bibinfo{author}{\bibfnamefont{A.}~\bibnamefont{Taranc\'on}},
  \bibinfo{journal}{Phys. Rev. B} \textbf{\bibinfo{volume}{67}},
  \bibinfo{pages}{214425} (\bibinfo{year}{2003}),
  \urlprefix\url{http://link.aps.org/doi/10.1103/PhysRevB.67.214425}.

\bibitem[{\citenamefont{Young and Katzgraber}(2004)}]{young:04}
\bibinfo{author}{\bibfnamefont{A.~P.} \bibnamefont{Young}} \bibnamefont{and}
  \bibinfo{author}{\bibfnamefont{H.~G.} \bibnamefont{Katzgraber}},
  \bibinfo{journal}{Phys. Rev. Lett.} \textbf{\bibinfo{volume}{93}},
  \bibinfo{pages}{207203} (\bibinfo{year}{2004}),
  \eprint{arXiv:cond-mat/0407031}.

\bibitem[{\citenamefont{Leuzzi et~al.}(2008)\citenamefont{Leuzzi, Parisi,
  Ricci-Tersenghi, and Ruiz-Lorenzo}}]{leuzzi:08}
\bibinfo{author}{\bibfnamefont{L.}~\bibnamefont{Leuzzi}},
  \bibinfo{author}{\bibfnamefont{G.}~\bibnamefont{Parisi}},
  \bibinfo{author}{\bibfnamefont{F.}~\bibnamefont{Ricci-Tersenghi}},
  \bibnamefont{and} \bibinfo{author}{\bibfnamefont{J.~J.}
  \bibnamefont{Ruiz-Lorenzo}}, \bibinfo{journal}{Phys. Rev. Lett.}
  \textbf{\bibinfo{volume}{101}}, \bibinfo{pages}{107203}
  (\bibinfo{year}{2008}),
  \urlprefix\url{http://link.aps.org/doi/10.1103/PhysRevLett.101.107203}.

\bibitem[{\citenamefont{Leuzzi et~al.}(2009)\citenamefont{Leuzzi, Parisi,
  Ricci-Tersenghi, and Ruiz-Lorenzo}}]{leuzzi:09}
\bibinfo{author}{\bibfnamefont{L.}~\bibnamefont{Leuzzi}},
  \bibinfo{author}{\bibfnamefont{G.}~\bibnamefont{Parisi}},
  \bibinfo{author}{\bibfnamefont{F.}~\bibnamefont{Ricci-Tersenghi}},
  \bibnamefont{and} \bibinfo{author}{\bibfnamefont{J.~J.}
  \bibnamefont{Ruiz-Lorenzo}}, \bibinfo{journal}{Phys. Rev. Lett.}
  \textbf{\bibinfo{volume}{103}}, \bibinfo{pages}{267201}
  (\bibinfo{year}{2009}), \eprint{arXiv:0811.3435}.

\bibitem[{\citenamefont{Leuzzi et~al.}(2011)\citenamefont{Leuzzi, Parisi,
  Ricci-Tersenghi, and Ruiz-Lorenzo}}]{leuzzi:11}
\bibinfo{author}{\bibfnamefont{L.}~\bibnamefont{Leuzzi}},
  \bibinfo{author}{\bibfnamefont{G.}~\bibnamefont{Parisi}},
  \bibinfo{author}{\bibfnamefont{F.}~\bibnamefont{Ricci-Tersenghi}},
  \bibnamefont{and}
  \bibinfo{author}{\bibfnamefont{J.}~\bibnamefont{Ruiz-Lorenzo}},
  \bibinfo{journal}{Philosophical Magazine} \textbf{\bibinfo{volume}{91}},
  \bibinfo{pages}{1917} (\bibinfo{year}{2011}),
  \eprint{https://doi.org/10.1080/14786435.2010.534741},
  \urlprefix\url{https://doi.org/10.1080/14786435.2010.534741}.

\bibitem[{\citenamefont{Ba\~{n}os et~al.}({2012})\citenamefont{Ba\~{n}os, Cruz,
  Fernandez, Gil-Narvion, Gordillo-Guerrero, Guidetti, Iniguez, Maiorano,
  Marinari, Mart\'{i}n-Mayor et~al.}}]{janus:12}
\bibinfo{author}{\bibfnamefont{R.~A.} \bibnamefont{Ba\~{n}os}},
  \bibinfo{author}{\bibfnamefont{A.}~\bibnamefont{Cruz}},
  \bibinfo{author}{\bibfnamefont{L.~A.} \bibnamefont{Fernandez}},
  \bibinfo{author}{\bibfnamefont{J.~M.} \bibnamefont{Gil-Narvion}},
  \bibinfo{author}{\bibfnamefont{A.}~\bibnamefont{Gordillo-Guerrero}},
  \bibinfo{author}{\bibfnamefont{M.}~\bibnamefont{Guidetti}},
  \bibinfo{author}{\bibfnamefont{D.}~\bibnamefont{Iniguez}},
  \bibinfo{author}{\bibfnamefont{A.}~\bibnamefont{Maiorano}},
  \bibinfo{author}{\bibfnamefont{E.}~\bibnamefont{Marinari}},
  \bibinfo{author}{\bibfnamefont{V.}~\bibnamefont{Mart\'{i}n-Mayor}},
  \bibnamefont{et~al.}, \bibinfo{journal}{Proc. Natl. Acad. Sci. USA}
  \textbf{\bibinfo{volume}{{109}}}, \bibinfo{pages}{6452}
  (\bibinfo{year}{{2012}}).

\bibitem[{\citenamefont{Larson et~al.}(2013)\citenamefont{Larson, Katzgraber,
  Moore, and Young}}]{larson:13}
\bibinfo{author}{\bibfnamefont{D.}~\bibnamefont{Larson}},
  \bibinfo{author}{\bibfnamefont{H.~G.} \bibnamefont{Katzgraber}},
  \bibinfo{author}{\bibfnamefont{M.~A.} \bibnamefont{Moore}}, \bibnamefont{and}
  \bibinfo{author}{\bibfnamefont{A.~P.} \bibnamefont{Young}},
  \bibinfo{journal}{Phys. Rev. B} \textbf{\bibinfo{volume}{87}},
  \bibinfo{pages}{024414} (\bibinfo{year}{2013}), \eprint{arXiv:1211.7297}.

\bibitem[{\citenamefont{Baity-Jesi
  et~al.}(2014{\natexlab{a}})\citenamefont{Baity-Jesi, Ba\~{n}os, Cruz,
  Fernandez, Gil-Narvion, Gordillo-Guerrero, Iniguez, Maiorano, F., Marinari
  et~al.}}]{janus:14b}
\bibinfo{author}{\bibfnamefont{M.}~\bibnamefont{Baity-Jesi}},
  \bibinfo{author}{\bibfnamefont{R.~A.} \bibnamefont{Ba\~{n}os}},
  \bibinfo{author}{\bibfnamefont{A.}~\bibnamefont{Cruz}},
  \bibinfo{author}{\bibfnamefont{L.~A.} \bibnamefont{Fernandez}},
  \bibinfo{author}{\bibfnamefont{J.~M.} \bibnamefont{Gil-Narvion}},
  \bibinfo{author}{\bibfnamefont{A.}~\bibnamefont{Gordillo-Guerrero}},
  \bibinfo{author}{\bibfnamefont{D.}~\bibnamefont{Iniguez}},
  \bibinfo{author}{\bibfnamefont{A.}~\bibnamefont{Maiorano}},
  \bibinfo{author}{\bibfnamefont{M.}~\bibnamefont{F.}},
  \bibinfo{author}{\bibfnamefont{E.}~\bibnamefont{Marinari}},
  \bibnamefont{et~al.}, \bibinfo{journal}{Phys. Rev. E}
  \textbf{\bibinfo{volume}{89}}, \bibinfo{pages}{032140}
  (\bibinfo{year}{2014}{\natexlab{a}}), \eprint{arXiv:1307.4998}.

\bibitem[{\citenamefont{Baity-Jesi
  et~al.}(2014{\natexlab{b}})\citenamefont{Baity-Jesi, Ba\~{n}os, Cruz,
  Fernandez, Gil-Narvion, Gordillo-Guerrero, Iniguez, Maiorano, F., Marinari
  et~al.}}]{janus:14c}
\bibinfo{author}{\bibfnamefont{M.}~\bibnamefont{Baity-Jesi}},
  \bibinfo{author}{\bibfnamefont{R.~A.} \bibnamefont{Ba\~{n}os}},
  \bibinfo{author}{\bibfnamefont{A.}~\bibnamefont{Cruz}},
  \bibinfo{author}{\bibfnamefont{L.~A.} \bibnamefont{Fernandez}},
  \bibinfo{author}{\bibfnamefont{J.~M.} \bibnamefont{Gil-Narvion}},
  \bibinfo{author}{\bibfnamefont{A.}~\bibnamefont{Gordillo-Guerrero}},
  \bibinfo{author}{\bibfnamefont{D.}~\bibnamefont{Iniguez}},
  \bibinfo{author}{\bibfnamefont{A.}~\bibnamefont{Maiorano}},
  \bibinfo{author}{\bibfnamefont{M.}~\bibnamefont{F.}},
  \bibinfo{author}{\bibfnamefont{E.}~\bibnamefont{Marinari}},
  \bibnamefont{et~al.}, \bibinfo{journal}{J. Stat. Mech.}
  \textbf{\bibinfo{volume}{2014}}, \bibinfo{pages}{P05014}
  (\bibinfo{year}{2014}{\natexlab{b}}), \eprint{arXiv:1403.2622}.

\bibitem[{\citenamefont{Takahashi et~al.}(2010)\citenamefont{Takahashi,
  Ricci-Tersenghi, and Kabashima}}]{takahashi2010finite}
\bibinfo{author}{\bibfnamefont{H.}~\bibnamefont{Takahashi}},
  \bibinfo{author}{\bibfnamefont{F.}~\bibnamefont{Ricci-Tersenghi}},
  \bibnamefont{and}
  \bibinfo{author}{\bibfnamefont{Y.}~\bibnamefont{Kabashima}},
  \bibinfo{journal}{Physical Review B} \textbf{\bibinfo{volume}{81}},
  \bibinfo{pages}{174407} (\bibinfo{year}{2010}).

\bibitem[{\citenamefont{M{\'e}zard and Parisi}(2001)}]{mezard:01}
\bibinfo{author}{\bibfnamefont{M.}~\bibnamefont{M{\'e}zard}} \bibnamefont{and}
  \bibinfo{author}{\bibfnamefont{G.}~\bibnamefont{Parisi}},
  \bibinfo{journal}{Eur. Phys. J. B} \textbf{\bibinfo{volume}{20}},
  \bibinfo{pages}{217} (\bibinfo{year}{2001}), \eprint{arXiv:cond-mat/0009418}.

\bibitem[{\citenamefont{Parisi and
  Ricci-Tersenghi}(2012)}]{parisi2012numerical}
\bibinfo{author}{\bibfnamefont{G.}~\bibnamefont{Parisi}} \bibnamefont{and}
  \bibinfo{author}{\bibfnamefont{F.}~\bibnamefont{Ricci-Tersenghi}},
  \bibinfo{journal}{Philosophical Magazine} \textbf{\bibinfo{volume}{92}},
  \bibinfo{pages}{341} (\bibinfo{year}{2012}).

\bibitem[{\citenamefont{Franz and Rieger}(1995)}]{franz:95}
\bibinfo{author}{\bibfnamefont{S.}~\bibnamefont{Franz}} \bibnamefont{and}
  \bibinfo{author}{\bibfnamefont{H.}~\bibnamefont{Rieger}},
  \bibinfo{journal}{Journal of Statistical Physics}
  \textbf{\bibinfo{volume}{79}}, \bibinfo{pages}{749} (\bibinfo{year}{1995}),
  ISSN \bibinfo{issn}{1572-9613},
  \urlprefix\url{http://dx.doi.org/10.1007/BF02184881}.

\bibitem[{\citenamefont{Franz et~al.}(1992)\citenamefont{Franz, Parisi, and
  Virasoro}}]{franz1992replica}
\bibinfo{author}{\bibfnamefont{S.}~\bibnamefont{Franz}},
  \bibinfo{author}{\bibfnamefont{G.}~\bibnamefont{Parisi}}, \bibnamefont{and}
  \bibinfo{author}{\bibfnamefont{M.~A.} \bibnamefont{Virasoro}},
  \bibinfo{journal}{Journal de Physique I} \textbf{\bibinfo{volume}{2}},
  \bibinfo{pages}{1869} (\bibinfo{year}{1992}).

\bibitem[{\citenamefont{Parisi}(1980)}]{parisi1980order}
\bibinfo{author}{\bibfnamefont{G.}~\bibnamefont{Parisi}},
  \bibinfo{journal}{Journal of Physics A: Mathematical and General}
  \textbf{\bibinfo{volume}{13}}, \bibinfo{pages}{1101} (\bibinfo{year}{1980}).

\bibitem[{\citenamefont{Kotliar et~al.}(1983)\citenamefont{Kotliar, Anderson,
  and Stein}}]{kotliar:83}
\bibinfo{author}{\bibfnamefont{G.}~\bibnamefont{Kotliar}},
  \bibinfo{author}{\bibfnamefont{P.~W.} \bibnamefont{Anderson}},
  \bibnamefont{and} \bibinfo{author}{\bibfnamefont{D.~L.} \bibnamefont{Stein}},
  \bibinfo{journal}{Phys. Rev. B} \textbf{\bibinfo{volume}{27}},
  \bibinfo{pages}{602} (\bibinfo{year}{1983}).

\bibitem[{\citenamefont{Leuzzi}(1999)}]{leuzzi:99}
\bibinfo{author}{\bibfnamefont{L.}~\bibnamefont{Leuzzi}},
  \bibinfo{journal}{Journal of Physics A: Mathematical and General}
  \textbf{\bibinfo{volume}{32}}, \bibinfo{pages}{1417} (\bibinfo{year}{1999}),
  \urlprefix\url{https://doi.org/10.1088%2F0305-4470%2F32%2F8%2F010}.

\bibitem[{\citenamefont{Leuzzi et~al.}(2015)\citenamefont{Leuzzi, Parisi,
  Ricci-Tersenghi, and Ruiz-Lorenzo}}]{leuzzi:15}
\bibinfo{author}{\bibfnamefont{L.}~\bibnamefont{Leuzzi}},
  \bibinfo{author}{\bibfnamefont{G.}~\bibnamefont{Parisi}},
  \bibinfo{author}{\bibfnamefont{F.}~\bibnamefont{Ricci-Tersenghi}},
  \bibnamefont{and} \bibinfo{author}{\bibfnamefont{J.~J.}
  \bibnamefont{Ruiz-Lorenzo}}, \bibinfo{journal}{Phys. Rev. B}
  \textbf{\bibinfo{volume}{91}}, \bibinfo{pages}{064202}
  (\bibinfo{year}{2015}),
  \urlprefix\url{https://link.aps.org/doi/10.1103/PhysRevB.91.064202}.

\bibitem[{\citenamefont{Hukushima and Nemoto}(1996)}]{hukushima:96}
\bibinfo{author}{\bibfnamefont{K.}~\bibnamefont{Hukushima}} \bibnamefont{and}
  \bibinfo{author}{\bibfnamefont{K.}~\bibnamefont{Nemoto}},
  \bibinfo{journal}{J. Phys. Soc. Japan} \textbf{\bibinfo{volume}{65}},
  \bibinfo{pages}{1604} (\bibinfo{year}{1996}),
  \eprint{arXiv:cond-mat/9512035}.

\bibitem[{\citenamefont{Marinari}(1998)}]{marinari:98b}
\bibinfo{author}{\bibfnamefont{E.}~\bibnamefont{Marinari}}, in
  \emph{\bibinfo{booktitle}{Advances in Computer Simulation}}, edited by
  \bibinfo{editor}{\bibfnamefont{J.}~\bibnamefont{Kerst\'esz}}
  \bibnamefont{and} \bibinfo{editor}{\bibfnamefont{I.}~\bibnamefont{Kondor}}
  (\bibinfo{publisher}{Springer-Verlag}, \bibinfo{year}{1998}).

\bibitem[{\citenamefont{de~Dominicis and Giardina}(2006)}]{dedominicis:06}
\bibinfo{author}{\bibfnamefont{C.}~\bibnamefont{de~Dominicis}}
  \bibnamefont{and} \bibinfo{author}{\bibfnamefont{I.}~\bibnamefont{Giardina}},
  \emph{\bibinfo{title}{{Random {F}ields and {S}pin {G}lasses}: a field theory
  approach}} (\bibinfo{publisher}{Cambridge University Press},
  \bibinfo{address}{Cambridge, England}, \bibinfo{year}{2006}).

\bibitem[{\citenamefont{Parisi and Ricci-Tersenghi}(2000)}]{parisi:00}
\bibinfo{author}{\bibfnamefont{G.}~\bibnamefont{Parisi}} \bibnamefont{and}
  \bibinfo{author}{\bibfnamefont{F.}~\bibnamefont{Ricci-Tersenghi}},
  \bibinfo{journal}{J. Phys. A: Math. Gen.} \textbf{\bibinfo{volume}{33}},
  \bibinfo{pages}{113} (\bibinfo{year}{2000}).

\bibitem[{\citenamefont{Marinari
  et~al.}(2000{\natexlab{b}})\citenamefont{Marinari, Parisi, Ricci-Tersenghi,
  Ruiz-Lorenzo, and Zuliani}}]{marinari:00}
\bibinfo{author}{\bibfnamefont{E.}~\bibnamefont{Marinari}},
  \bibinfo{author}{\bibfnamefont{G.}~\bibnamefont{Parisi}},
  \bibinfo{author}{\bibfnamefont{F.}~\bibnamefont{Ricci-Tersenghi}},
  \bibinfo{author}{\bibfnamefont{J.~J.} \bibnamefont{Ruiz-Lorenzo}},
  \bibnamefont{and} \bibinfo{author}{\bibfnamefont{F.}~\bibnamefont{Zuliani}},
  \bibinfo{journal}{J. Stat. Phys.} \textbf{\bibinfo{volume}{98}},
  \bibinfo{pages}{973} (\bibinfo{year}{2000}{\natexlab{b}}),
  \eprint{arXiv:cond-mat/9906076}.

\bibitem[{\citenamefont{Leuzzi and Parisi}(2013)}]{leuzzi:13b}
\bibinfo{author}{\bibfnamefont{L.}~\bibnamefont{Leuzzi}} \bibnamefont{and}
  \bibinfo{author}{\bibfnamefont{G.}~\bibnamefont{Parisi}},
  \bibinfo{journal}{Phys. Rev. B} \textbf{\bibinfo{volume}{88}},
  \bibinfo{pages}{224204} (\bibinfo{year}{2013}),
  \urlprefix\url{https://link.aps.org/doi/10.1103/PhysRevB.88.224204}.

\end{thebibliography}

\end{document}